\documentclass[lettersize,journal]{IEEEtran}

\usepackage{amsmath,amssymb,amsfonts}
\usepackage{algorithmic}
\usepackage{algorithm}
\usepackage{array}
\usepackage[caption=false,font=normalsize,labelfont=bf,textfont=rm]{subfig}
\usepackage{textcomp}
\usepackage{stfloats}
\usepackage{url}
\usepackage{verbatim}
\usepackage{graphicx}
\usepackage{siunitx}
\usepackage{cite}
\usepackage{capt-of}
\usepackage{booktabs}
\usepackage{hyperref}

\usepackage{makecell}
\usepackage[table]{xcolor}

    \def\Complex{{\rm\rule[.23ex]{.03em}{1.1ex}\kern-.3em{C}}}

    \newcommand{\be}{\begin{equation}} \newcommand{\ee}{\end{equation}}
    \newcommand{\bea}{\begin{eqnarray}} \newcommand{\eea}{\end{eqnarray}}
    \newcommand{\benum}{\begin{enumerate}} \newcommand{\eenum}{\end{enumerate}}

    \usepackage{xcolor}

\usepackage[english]{babel}
\newtheorem{theorem}{Theorem}
\newtheorem{corollary}{Corollary}
\newtheorem{remark}{Remark}

\begin{document}

\title{CommUNext: Deep Learning-Based Cross-Band and Multi-Directional Signal Prediction}

\author{Chi-Jui Sung,
        Fan-Hao Lin,
        Tzu-Hao Huang,
        Chu-Hsiang Huang,~\IEEEmembership{Member,~IEEE,}
        Hui Chen,~\IEEEmembership{Member,~IEEE,}
        Chao-Kai Wen,~\IEEEmembership{Fellow,~IEEE,}
        Henk Wymeersch,~\IEEEmembership{Fellow,~IEEE}

\thanks{The work of C. J. Sung, F. H. Lin, T. H. Huang, and C. K. Wen was supported in part by the National Science and Technology Council (NSTC) of Taiwan under Grants NSTC 114-2221-E-110-031-MY3 and NSTC 114-2218-E-110-005, and in part by the Sixth Generation Communication and Sensing Research Center, funded by the Higher Education SPROUT Project of the Ministry of Education (MOE) of Taiwan.
The work of C. H. Huang was supported in part by the NSTC of Taiwan under Grant NSTC 113-2222-E-002-010.
The work of H. Chen and H. Wymeersch was supported in part by the SNS JU project 6G-DISAC under the EU Horizon Europe research and innovation programme under Grant Agreement No. 101139130, and in part by HORIZON-MSCA through Project SMALL-6G under Grant 101201808. \emph{(Corresponding author: Chao-Kai Wen.)}}

\thanks{Chi-Jui Sung was with the Institute of Communications Engineering, National Sun Yat-sen University, Kaohsiung 80424, Taiwan, and is now with Tron Future Tech Inc (e-mail: johnnysung28@gmail.com).}

\thanks{Fan-Hao Lin and Chao-Kai Wen are with the Institute of Communications Engineering, National Sun Yat-sen University, Kaohsiung 80424, Taiwan (e-mail: 22135lin@gmail.com; chaokai.wen@mail.nsysu.edu.tw).}

\thanks{Tzu-Hao Huang is with the Institute of Electrical Control Engineering, National Yang Ming Chiao Tung University, Hsinchu 300, Taiwan (e-mail: peter94135@gmail.com).}

\thanks{Chu-Hsiang Huang is with the Department of Electrical Engineering, National Taiwan University, Taipei 10617, Taiwan (e-mail: chuhsianh@ntu.edu.tw).}

\thanks{Hui Chen and Henk Wymeersch are with the Department of Electrical Engineering, Chalmers University of Technology, 41296 Gothenburg, Sweden (e-mail: hui.chen@chalmers.se; henkw@chalmers.se). Hui Chen is also with the Department of Electronic and Electrical Engineering, University College London, London WC1E 7JE, U.K. (e-mail: hui.chen@ucl.ac.uk).} 
}

\markboth{IEEE Transactions on Wireless Communications, VOL. , NO. , MARCH 2026}%
{Shell \MakeLowercase{\textit{et al.}}: CommUNext: Deep Learning-Based Cross-Band and Multi-Directional Signal Reconstruction and Prediction}


\maketitle

\begin{abstract}
Sixth-generation (6G) networks are envisioned to achieve full-band cognition by jointly utilizing spectrum resources from Frequency Range~1 (FR1) to Frequency Range~3 (FR3, 7--24\,GHz). Realizing this vision faces two challenges. First, physics-based ray tracing (RT), the standard tool for network planning and coverage modeling, becomes computationally prohibitive for multi-band and multi-directional analysis over large areas. Second, current 5G systems rely on inter-frequency measurement gaps for carrier aggregation and beam management, which reduce throughput, increase latency, and scale poorly as bands and beams proliferate. These limitations motivate a data-driven approach to infer high-frequency characteristics from low-frequency observations.
This work proposes CommUNext, a unified deep learning framework for cross-band, multi-directional signal strength (SS) prediction. The framework leverages low-frequency coverage data and crowd-aided partial measurements at the target band to generate high-fidelity FR3 predictions. Two complementary architectures are introduced: Full CommUNext, which substitutes costly RT simulations for large-scale offline modeling, and Partial CommUNext, which reconstructs incomplete low-frequency maps to mitigate measurement gaps in real-time operation.
Experimental results show that CommUNext delivers accurate and robust high-frequency SS prediction even with sparse supervision, substantially reducing both simulation and measurement overhead.
\end{abstract}


\section{Introduction}
\IEEEPARstart{S}{ixth}-generation (6G) networks aim to integrate the full spectrum from Frequency Range 1 (FR1) to Frequency Range 3 (FR3), including the 7--24 GHz upper mid band~\cite{kang2024upper}, into a unified spectrum resource pool to support the six IMT-2030 usage scenarios~\cite{itu2023imt2030framework}. Achieving this vision requires \emph{full-band cognition}, which is the capability to sense and predict channel conditions across frequencies, beam directions, spatial locations, and time while minimizing active probing and control overhead. Such cognition is essential for cross-band resource optimization, carrier aggregation (CA), and beam management in large-scale and dynamic deployments.

In the early stage of 6G network design, physics-based \emph{ray tracing} (RT) serves as a vital tool for site planning, coverage evaluation, and link-budget analysis. RT simulations account for material properties, reflections, diffractions, and scattering to generate frequency-specific and direction-aware coverage maps. However, each simulation run typically produces only a single-band, single-beam output (e.g., Sionna~\cite{sionna2025} and Remcom Wireless InSite~\cite{WirelessInSite}). Extending RT to multi-band and multi-directional analyzes over large urban areas drastically increases computational complexity and memory usage, often requiring hours or even days per region.\footnote{%
For example, generating 133{,}956 received signal strength maps using Sionna, as in this work, requires approximately one month on a personal computer equipped with an RTX 2080 Ti GPU.}
Although RT is primarily designed for offline coverage modeling rather than real-time operation, its extensive computational cost still poses a significant efficiency bottleneck even for offline analyzes.

For real-time operation, current 5G New Radio (NR) systems rely on inter-frequency \emph{measurement gaps} to discover carriers for CA, execute inter-band handovers, and perform beam management. During each gap, user equipment (UE) suspends reception on the serving link to probe another frequency (e.g., every 20/40/80\,ms with a multi-slot gap)~\cite{3gpp38331,3gpp38133}. While this mechanism is workable in existing systems, it reduces throughput, increases latency and power consumption, and scales poorly as 6G extends to full spectrum with finer beam codebooks. The cumulative cost of frequent multi-band probing renders gap-based coordination unsustainable.

{
These offline and operational challenges share a common origin in the evolving 6G spectrum landscape. In future deployments, the FR3 spectrum is expected to be co-located with existing FR1 macro sites, enabling infrastructure reuse and preserving large scale spatial correlation across frequency bands~\cite{5GAmericas2024PurposeBackground,Ericsson2025ReuseGrid}. This co-location implies that abundant and stable low frequency telemetry, such as 3.5\,GHz, can serve as a spatial proxy for higher frequency environments.} However, in practical operation, such low frequency data may be incomplete, as user reports can be sparse, delayed, or unevenly distributed across regions. Moreover, sparse high band samples may be available rather than entirely absent. These observations motivate a data driven approach that exploits rich but occasionally incomplete low band data together with sparse high band samples to predict multi directional signal strength (SS) maps, thereby enabling scalable full band cognition while reducing reliance on computationally intensive RT and extensive measurement campaigns.

Building on this intuition, we propose CommUNext, a deep learning-based framework for \emph{cross-band} and \emph{multi-directional} signal prediction. The framework comprises two complementary architectures tailored to different data availability regimes. When a complete 3.5\,GHz SS coverage map is available, Full CommUNext leverages it as a strong reference to predict high-band SS maps, addressing the limitations of RT-based offline modeling. When the low-frequency data are partially observed, such as when cloud-aggregated measurements are sparse or regionally missing, Partial CommUNext reconstructs the missing low-frequency map and transfers latent cross-band knowledge to enhance high-band prediction, thereby mitigating the impact of measurement gaps in live operation. In both cases, the framework integrates crowd-aided context by aggregating geo-aligned user reports that may include partial (opportunistic) measurements at the target band within the relevant time period, capturing dynamic blockage and environmental variations.\footnote{The operational mechanisms of the crowd-aided setting are detailed in Appendix~\ref{sec:appendix_a}.}

The main contributions of this work are summarized as follows:
\begin{itemize}
    \item Proposed a unified data-driven framework that predicts \emph{high-frequency, multi-directional} SS maps from low-frequency (3.5\,GHz) coverage data and sparse high-frequency samples. The framework enables cross-band and multi-directional inference under both complete and incomplete reference-band conditions, thereby significantly reducing dependence on computationally intensive RT and measurement campaigns.

    \item Designed two complementary network architectures, Full CommUNext and Partial CommUNext, to realize this framework. The former leverages complete 3.5\,GHz coverage maps as strong priors for cross-band prediction, while the latter reconstructs missing low-frequency information and transfers latent knowledge directly to enhance prediction under sparse supervision. Partial CommUNext achieves higher computational efficiency than naively passing the reconstructed 3.5\,GHz coverage maps again to Full CommUNext.

    \item Conducted extensive evaluations using RT-based datasets to validate the proposed framework under various input prior conditions. The evaluations and analyses provide insights into which input priors are most relevant to cross-band prediction.

\end{itemize}
The remainder of this paper is organized as follows.
Section~II reviews related works; Section~III describes data generation and preprocessing; Section~IV details the data-driven framework and CommUNext; Section~V presents experimental results; and Section~VI concludes.

\section{Related Works}

Traditional path-loss estimation relies on deterministic RT, which models reflection, diffraction, and scattering. While accurate in static environments, these methods are computationally expensive and unsuitable for large-scale or real-time tasks. Deep learning-based frameworks \cite{li2025generalized,ma2025dualphase,liu2025pinn} address this issue by offering fast and scalable inference with high accuracy. Early contributions include \cite{levie2021radiounet,li2024geo2sigmap,qiu2022pseudort}, and a survey is provided in \cite{yapar2024plchallenge}. These works collectively establish deep learning as a practical alternative to purely physics-based methods for radio propagation modeling.

Various architectures have been explored for radio map prediction. U-Net and its variants dominate early designs due to their strong spatial reconstruction capability \cite{li2024geo2sigmap,li2025generalized}. Attention U-Nets improve spatial feature weighting, enabling joint throughput and link-quality prediction \cite{lin2025geo2commap}. Beyond CNNs, GNN-based methods capture irregular spatial-spectral relations \cite{li2024radiogat}, while GAN-based approaches (e.g., \cite{zhou2025tiregan}) refine predictions through adversarial and task-driven learning. These diverse architectures highlight the importance of inductive biases in capturing propagation characteristics across heterogeneous environments.

Recent studies have further extended these methods toward cross-band prediction. HORCRUX focuses on frequency-division duplex systems by predicting downlink channels from uplink observations, emphasizing cross-environment generalization and scalability \cite{banerjee2024horcrux}. This task is link-level and relies on per-link uplink measurements, whereas our setting targets region-level, multi-directional SS maps driven by low-frequency coverage and sparse high-band samples. Another line of work integrates simulation and digital-twin modeling. RadioTwin combines material property modeling with RT engines to enable cross-band channel prediction \cite{an2025radiotwin}. While this approach embeds geometry and materials explicitly, it requires high-fidelity modeling and incurs substantial simulation cost. In contrast, our framework adopts a data-driven paradigm that leverages low-frequency maps and sparse high-band samples to reduce dependence on heavy RT and large-scale field surveys.

A complementary direction estimates system-level metrics. Geo2ComMap predicts higher-layer communication metrics such as throughput, Channel Quality Indicator (CQI), and Rank Indicator (RI), providing region-level performance maps \cite{lin2025geo2commap}. Transformer-based methods further exploit historical multi-band measurements to forecast achievable rates over time \cite{chen2025transformerRatePrediction}. These works operate at the service or system layer and primarily leverage performance indicators or temporal sequences, whereas our framework focuses on physical-layer SS maps with explicit multi-directional structure and cross-band transfer. Table~\ref{tab:comparison_methods} summarizes the key differences among representative works.

Datasets have also driven progress in this field. DeepMIMO \cite{alkhateeb2019deepmimo} provides standardized mmWave MIMO data, while RadioMapSeer \cite{yapar2022dataset} supplies urban path-loss layers. The Indoor Radio Map Dataset \cite{bakirtzis2024indoordataport} introduces large-scale benchmarks with sparse supervision, and \cite{lin2025geo2commap} adds datasets covering throughput, CQI, and RI. These datasets reflect the growing demand for multi-band, multi-metric benchmarks that bridge physical- and system-level perspectives.

\begin{table*}[t]
\centering
{
\caption{Comparison with Representative Learning-Based Radio Map Prediction Approaches}
\label{tab:comparison_methods}
\begin{tabular}{l l l l l}
\hline
\textbf{Method}
& \textbf{Prediction Scale}
& \textbf{Prediction Target}
& \textbf{Prediction Domain}
& \textbf{Input Priors} \\
\hline
\addlinespace
HORCRUX~\cite{banerjee2024horcrux}
& Link-level
& Downlink channel
& \makecell[l]{Specific \\ (Single-band, single-link)}
& \makecell[l]{Heavy measurement cost \\ (Per-link uplink measurements)} \\
\addlinespace
RadioTwin~\cite{an2025radiotwin}
& Link-level
& Material radio parameters
&  \makecell[l]{Highly general \\ (Cross-band, cross-link)}
& \makecell[l]{High complexity \\ (3D scene mesh and \\ physics-based RT paths)} \\
\addlinespace
Geo2ComMap~\cite{lin2025geo2commap}
& Region-level
& Communication performance
&  \makecell[l]{Limited \\ (Single-band, region-level)}
& \makecell[l]{Moderate complexity \\ (Sparse region-level \\ performance maps)} \\
\addlinespace
\rowcolor{gray!15}
CommUNext
& Region-level
& Multi-directional SS maps
&  \makecell[l]{General \\ (Cross-band, region-level)}
& \makecell[l]{Low cost \\ (3.5~GHz coverage maps \\ and sparse high-band measurements)} \\
\hline
\addlinespace
\end{tabular}
}
\end{table*}

\section{Data Generation and Preprocessing} \label{sec:DataGen}
This section describes the data generation and preprocessing pipeline employed in this work. The process encompasses the acquisition of geographical information, simulation methodology, parameter configuration, and the preparation of both input features and ground-truth labels for the data-driven neural network described in Section~\ref{sec:CommUNext}.

\subsection{Automated Data Generation Toolchain}
\begin{figure}[!t]
    \centering
    \includegraphics[width=\columnwidth]{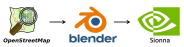}
    \caption{Automated data generation toolchain integrating OSM, Blender, and Sionna.}
    \label{fig:toolchain}
\end{figure}

To emulate multi-directional and multi-band propagation, we developed an automated toolchain inspired by Geo2SigMap \cite{li2024geo2sigmap}. The framework integrates geographic information retrieval, 3D building modeling, and RT simulation, enabling the construction of realistic wireless datasets from real urban environments. The overall workflow is shown in Fig.~\ref{fig:toolchain} and involves three open-source components: OpenStreetMap (OSM), Blender, and NVIDIA Sionna.

\begin{figure}[!t]
    \centering
    \includegraphics[width=\columnwidth]{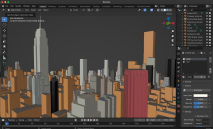}
    \caption{Example of importing OSM building data into Blender.}
    \label{fig:blender}
\end{figure}

\subsubsection{OSM and Blender}
Geographic data were obtained from OSM for the New York City metropolitan area covering $[40.477399, 40.917577]$ latitude and $[-74.259090, -73.700272]$ longitude. The area was partitioned into \SI{500}{m}$\times$\SI{500}{m} blocks, and those with more than 20\% building coverage were retained. Using the \texttt{blosm} plugin, building footprints were imported into Blender, converted into 3D meshes, and assigned material properties according to~\cite{itu2023p2040}. Each block produced (i) a building height map, (ii) Mitsuba-compatible 3D scene files, and (iii) metadata including boundaries, coverage ratio, and a Universally Unique Identifier (UUID). Fig.~\ref{fig:blender} illustrates an example of OSM data imported into Blender.

\subsubsection{NVIDIA Sionna}
The 3D scenes were simulated using the \textit{NVIDIA Sionna Neural Radio Framework} \cite{sionna2025}, a GPU-accelerated library with differentiable RT \cite{hoydis2023sionnart,aoudia2025sionnarttr}. Sionna models reflection, diffraction, and scattering, and outputs path gain, received signal strength (RSS), and channel impulse responses.\footnote{Channel characterization based on radio maps and the relevance of FR1-FR3 prediction are provided in Appendix~\ref{app:channel_rmap}.}  In this work, it was configured to:
\begin{itemize}
    \item Perform multi-band simulations at 3.5 and 7\,GHz with user-specified transmitters, antenna patterns, and beam orientations.
    \item Generate omnidirectional coverage maps and eight-direction SS maps as supervised training labels.
\end{itemize}

\subsection{Data Generation}
To create datasets for model training, we combined OSM-derived building maps with NVIDIA Sionna RT simulations at 3.5 and 7\,GHz. The workflow produces three key outputs: (i) building height maps $\mathbf{B}$, (ii) 3.5\,GHz coverage maps $\mathbf{S}_c$, and (iii) eight-direction 7\,GHz SS maps $\{\mathbf{S}_{d1}, \mathbf{S}_{d2}, \dots, \mathbf{S}_{d8}\}$, all aligned at a unified resolution of $128 \times 128$ pixels.

\subsubsection{Building Maps}
For each \SI{500}{m}$\times$\SI{500}{m} block in the New York City area, building footprints from OSM are converted into 3D meshes using Blender. Each block produces a building height map represented as a $512 \times 512$ resolution patch. Each patch is then downsampled to a $128 \times 128$ building height map $\mathbf{B}$ to match the resolution of the signal coverage maps, as illustrated in Fig.~\ref{fig:building_coverage}(a).

\subsubsection{3.5\,GHz Coverage Maps}
\begin{figure}[!t]
\centering
\captionsetup[subfloat]{labelformat=empty}
\subfloat[\footnotesize{(a) Building Map (m)}]
{\includegraphics[width=0.5\columnwidth]{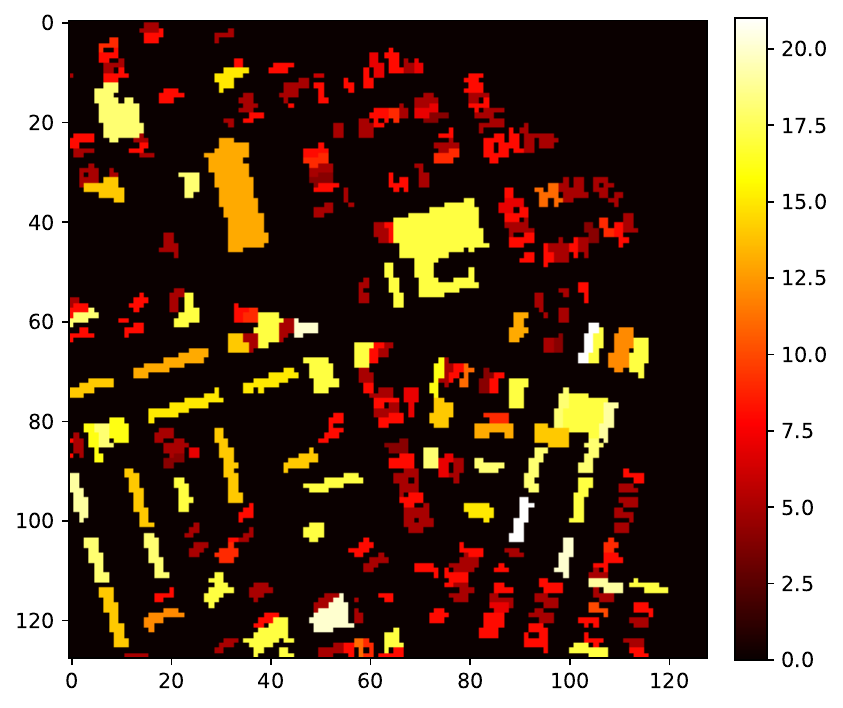}}\hfil
\subfloat[\footnotesize{(b) 3.5~GHz Coverage Map (dBm)}]
{\includegraphics[width=0.5\columnwidth]{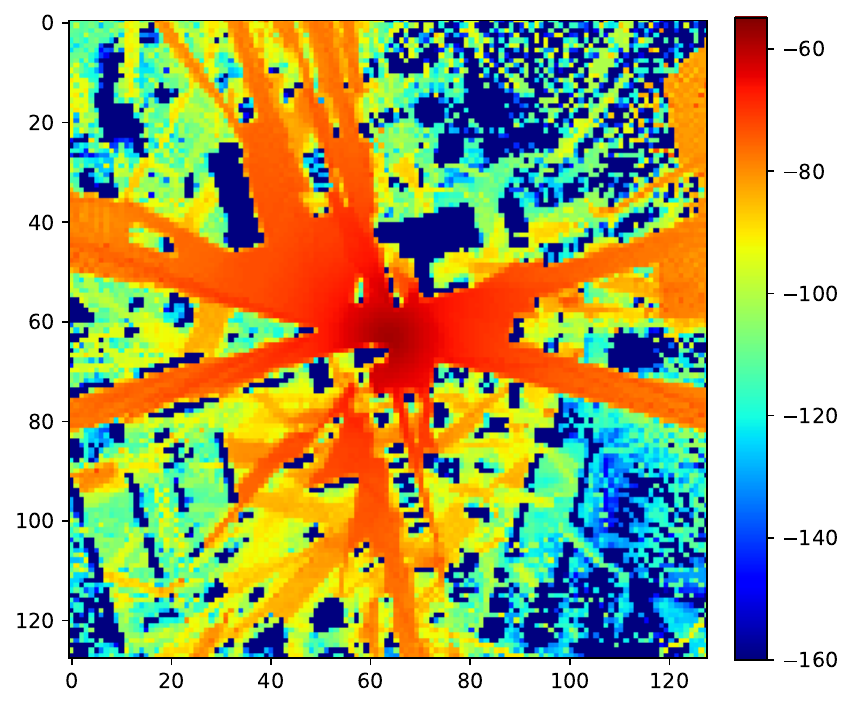}}
\caption{Example of (a) a building map $\mathbf{B}$ and (b) the corresponding 3.5\,GHz coverage map $\mathbf{S}_c$.}
\label{fig:building_coverage}
\end{figure}

Using Sionna’s ray tracer, 3.5\,GHz propagation is simulated with isotropic base station transmitter antennas located at the center of each building map. The simulator accounts for line-of-sight (LoS) paths, reflections, and diffractions, producing RSS values in dBm on a $128 \times 128$ grid $\mathbf{S}_c$, as shown in Fig.~\ref{fig:building_coverage}(b). Table~\ref{tab:sim_params} summarizes the key simulation parameters, including carrier frequency, antenna height, and spatial resolution.

\begin{table}[!t]
\centering
\caption{Simulation parameters for generating 3.5\,GHz and 7\,GHz maps in Sionna. Parameters not listed under 7\,GHz are identical to those used for 3.5\,GHz.}
\label{tab:sim_params}
\begin{tabular}{l l}
\hline
\textbf{Parameter} & \textbf{Value} \\
\hline
Carrier frequency & 3.5\,GHz / 7\,GHz \\
BS antenna height & Max building height + 5\,m \\
BS location & Center of the building map \\
UE antenna height & 1.5\,m \\
Number of rays & $3\times10^{6}$ \\
\rowcolor{gray!15}
BS antenna gain & 0\,dBi (3.5\,GHz) / 6\,dBi (7\,GHz) \\
Reflection / Diffraction & Enabled / Enabled \\
Max reflection / diffraction order & 8 \\
\rowcolor{gray!15}
Antenna pattern & \makecell[l]{Isotropic (3.5\,GHz) /\\ Directional (TR38.901)\cite{3gpp38901r16} (7\,GHz)} \\
Antenna polarization & Dual-polarized (VH) \\
\rowcolor{gray!15}
Number of directions (7\,GHz only) & 8 ($0^{\circ},45^{\circ},\dots,315^{\circ}$) \\
Simulation area size & 512\,m $\times$ 512\,m \\
Building wall material & ITU \cite{ITU2015} brick \\
Building roof material & ITU \cite{ITU2015} plaster board \\
Spatial resolution & 4\,m/pixel \\
Output resolution & $128 \times 128$ \\
\hline
\end{tabular}
\end{table}

\subsubsection{7\,GHz Multi-Directional SS Maps}
\begin{figure}[!t]
\centering
\captionsetup[subfloat]{labelformat=empty}
\subfloat[\footnotesize{(a) 0$^{\circ}$}]
{\includegraphics[width=0.2\textwidth]{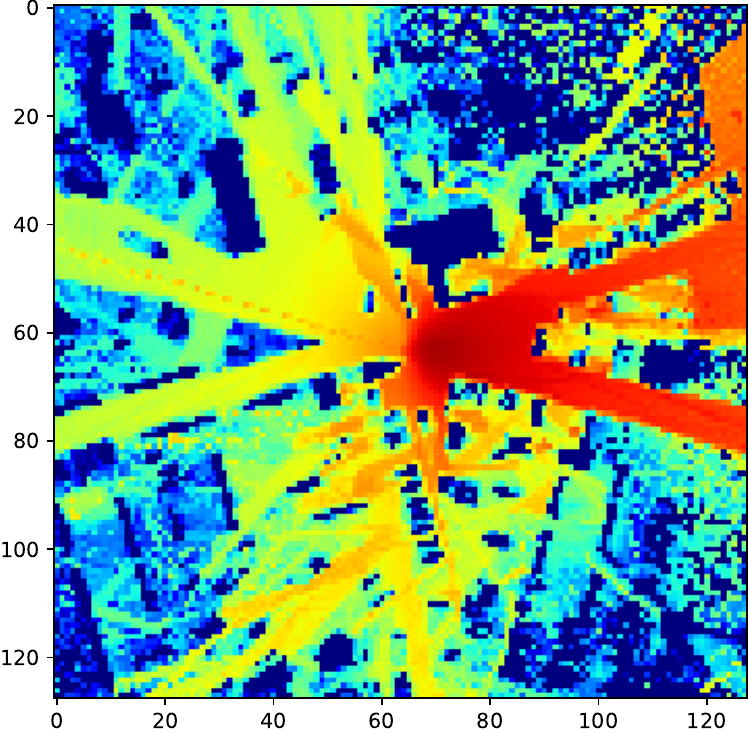}}\hfil
\subfloat[\footnotesize{(b) 45$^{\circ}$}]
{\includegraphics[width=0.2\textwidth]{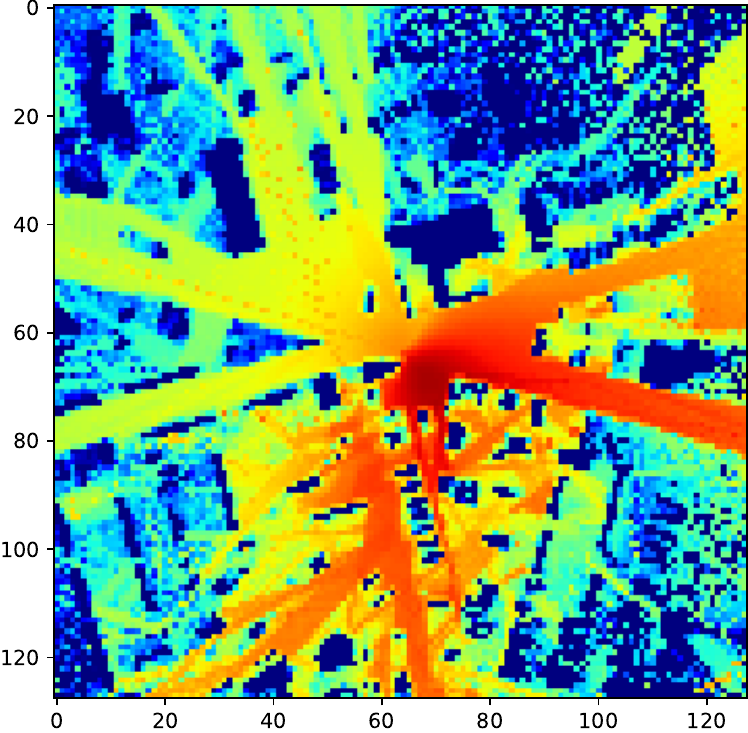}}\hfil
\subfloat[\footnotesize{(c) 90$^{\circ}$}]
{\includegraphics[width=0.2\textwidth]{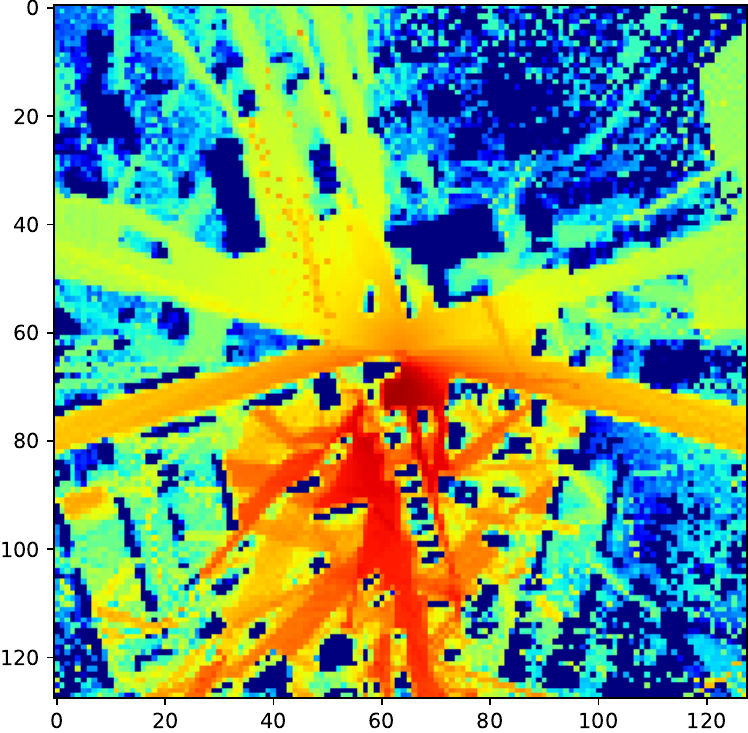}}\hfil
\subfloat[\footnotesize{(d) 135$^{\circ}$}]
{\includegraphics[width=0.2\textwidth]{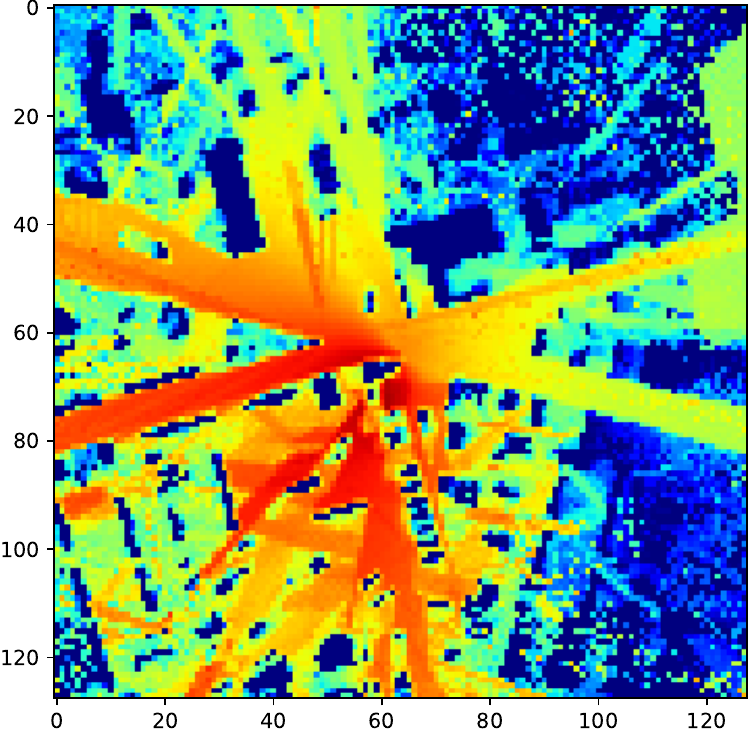}}\\

\subfloat[\footnotesize{(e) 180$^{\circ}$}]
{\includegraphics[width=0.2\textwidth]{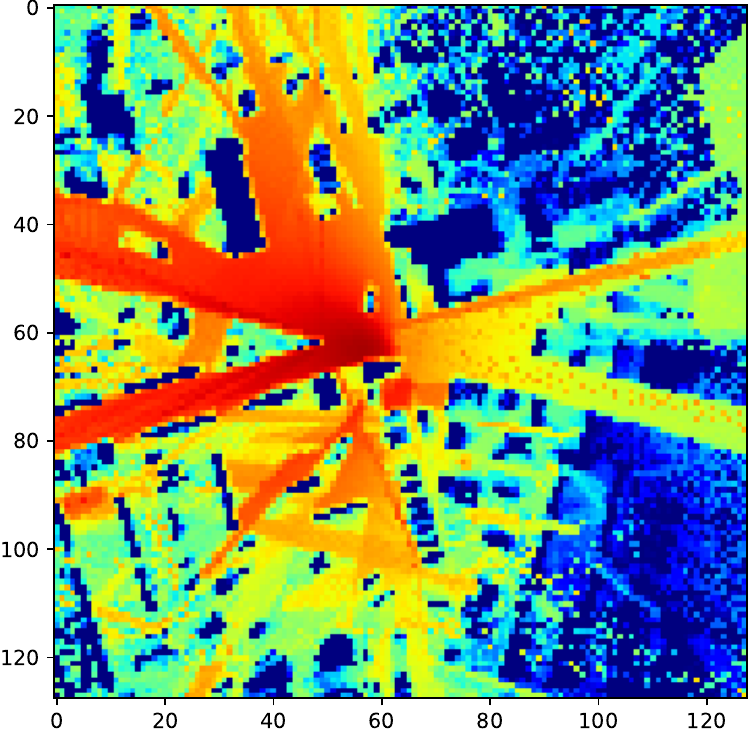}}\hfil
\subfloat[\footnotesize{(f) 225$^{\circ}$}]
{\includegraphics[width=0.2\textwidth]{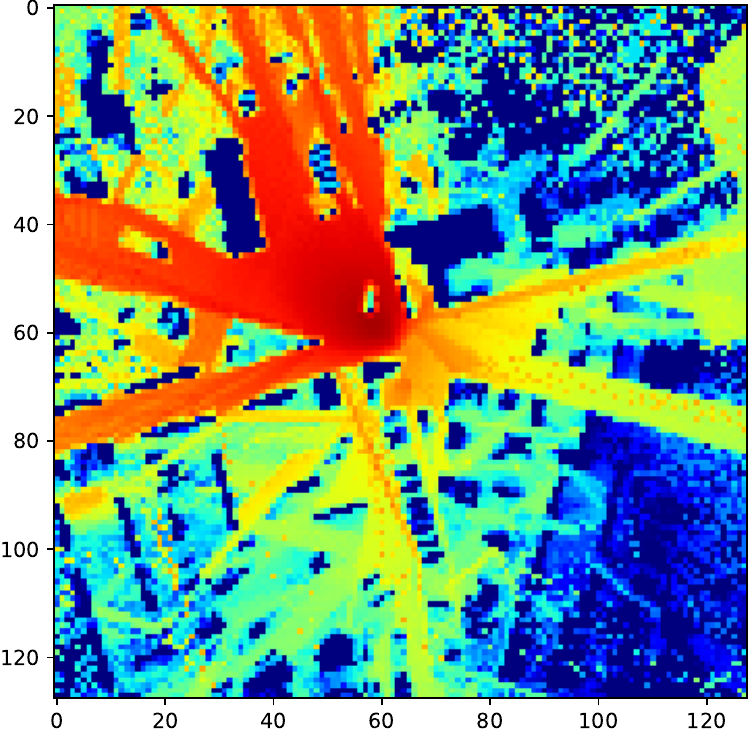}}\hfil
\subfloat[\footnotesize{(g) 270$^{\circ}$}]
{\includegraphics[width=0.2\textwidth]{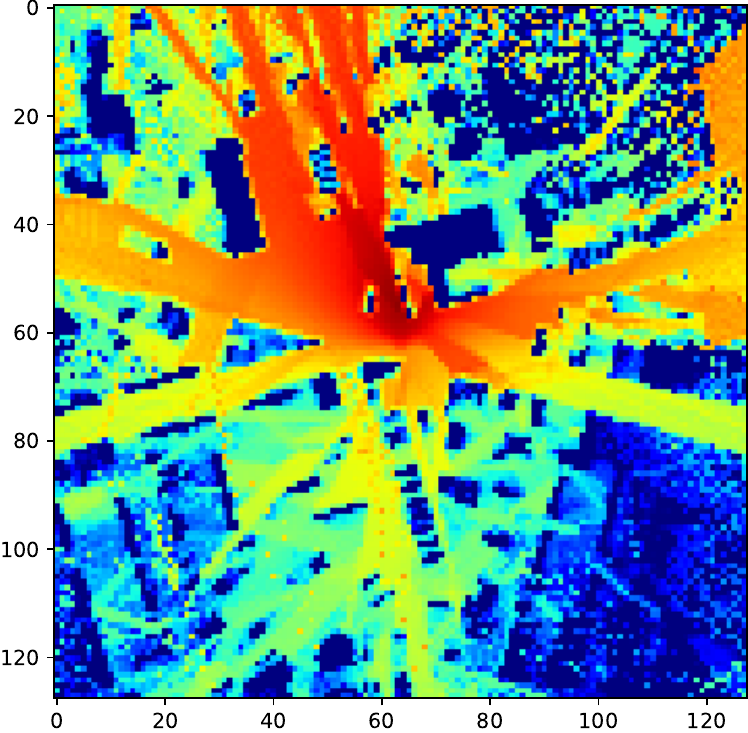}}\hfil
\subfloat[\footnotesize{(h) 315$^{\circ}$}]
{\includegraphics[width=0.2\textwidth]{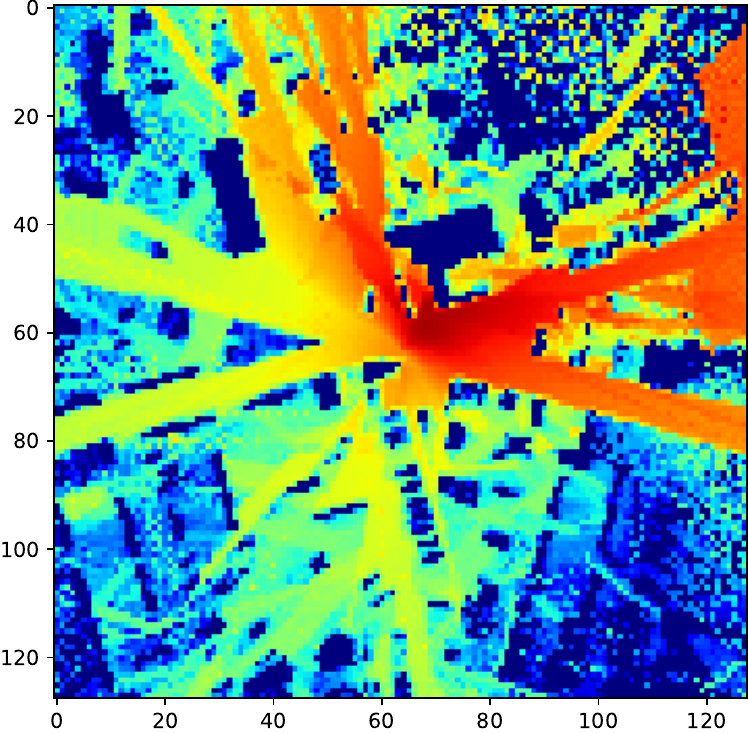}}\\
\vspace{0.25cm}
\includegraphics[width=0.40\textwidth]{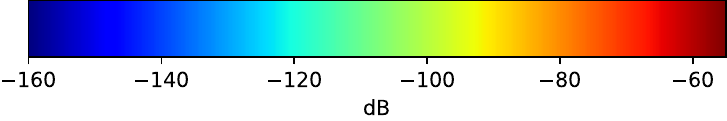}
\vspace{-0.25cm}
\caption{Eight generated 7\,GHz SS maps $\{\mathbf{S}_{d1}, \mathbf{S}_{d2}, \dots, \mathbf{S}_{d8}\}$, showing beam concentration along pointing directions and variations caused by blockages.}
\label{fig:sS_maps}
\end{figure}

We assume that the 7,GHz site is co-located with the 3.5,GHz site.\footnote{In practical deployments, slight antenna reorientation or displacement does not affect the results, since the proposed design incorporates sparse 7,GHz SS samples as reference inputs. The effectiveness of calibration using sparse samples has been demonstrated in~\cite{li2024geo2sigmap}.}
At 7\,GHz, Sionna is configured with TR38.901 directional antenna patterns oriented at eight principal angles $\{0^{\circ}, 45^{\circ}, \dots, 315^{\circ}\}$. For each transmitter, eight SS maps are generated, representing beamformed coverage in different directions. Fig.~\ref{fig:sS_maps} shows an example set of directional maps $\{\mathbf{S}_{d1}, \mathbf{S}_{d2}, \dots, \mathbf{S}_{d8}\}$, illustrating beam concentration along the pointing angles and variations caused by blockages. Table~\ref{tab:sim_params} highlights the parameters that differ from the 3.5\,GHz configuration.

\subsection{Data Preprocessing} \label{sec:dataset_preprocess}
To prepare the dataset for evaluation, several preprocessing steps were applied to ensure spatial consistency and representativeness of real-world measurement conditions, including dataset partitioning, sparse sampling of 7\,GHz maps, generation of LoS and non-line-of-sight (NLoS) masks, and synthesis of partially observed 3.5\,GHz coverage maps.

\subsubsection{Dataset Partitioning} \label{sec:data_partitioning}

A total of 14,884 \emph{non-overlapping} \SI{500}{m}$\times$\SI{500}{m} building height maps were obtained from OSM and Blender modeling. For each valid sample, one building map $\mathbf{B}$, one 3.5\,GHz coverage map $\mathbf{S}_c$, and eight directional 7\,GHz SS maps ${\mathbf{S}_{d1}, \dots, \mathbf{S}_{d8}}$ were required. Samples missing any of these components were discarded. The dataset was then divided into training, validation, and test sets with a ratio of 7:2:1 to ensure independence. To enable robust evaluation, this division was performed at the building map level, ensuring that no overlap exists among the building maps used for training, validation, and testing. 

\subsubsection{Sparse 7\,GHz SS Maps}\label{sec:7GHzSSMapGen}
To emulate limited real-world measurements, sparse 7\,GHz SS maps $\{\widetilde{\mathbf{S}}_{d1}, \dots, \widetilde{\mathbf{S}}_{d8}\}$ were derived from the ground-truth $\{\mathbf{S}_{d1}, \dots, \mathbf{S}_{d8}\}$ using two sampling strategies:
\begin{itemize}
\item \textit{Random sampling:} $N_d$ out of $128 \times 128$ pixels were randomly selected, with the remaining pixels set to $-160$\,dBm.

\item \textit{NLoS-guided sampling:} Similar to the random sampling strategy, $N_d$ pixels were selected, where a fraction $\gamma$ was drawn from NLoS regions and $(1-\gamma)$ from LoS regions, prioritizing blocked areas.
\end{itemize}
Both strategies retained a resolution of $128 \times 128$, producing eight sparse maps per sample.

\subsubsection{LoS and NLoS Masks}
To better guide the model in learning blocked regions, LoS and NLoS masks, denoted as $\mathbf{M}_\text{LoS}$ and $\mathbf{M}_\text{NLoS}$, were designed.
The LoS mask $\mathbf{M}_\text{LoS}$ is defined as a $128 \times 128$ binary map with values of 0 or 1. It is computed using the Bresenham line algorithm. For each pixel $(x_1, y_1)$ relative to a Tx at $(x_0, y_0)$, the connecting line is expressed as
\begin{equation}
y - y_0 = \frac{y_1 - y_0}{x_1 - x_0}(x - x_0),
\label{eq:line}
\end{equation}
and discretized into grid cells. If any cell (excluding the Tx) intersects a building, i.e., $[\mathbf{B}]_{x,y} > 0$, the pixel is marked as blocked; otherwise, it is classified as LoS.

\begin{figure}[!t]
\centering
{
\begin{minipage}{0.48\textwidth}
    \centering
    \captionsetup[subfloat]{labelformat=empty}
    \subfloat[\footnotesize{(a) LoS (green)}]
    {\includegraphics[width=0.45\textwidth]{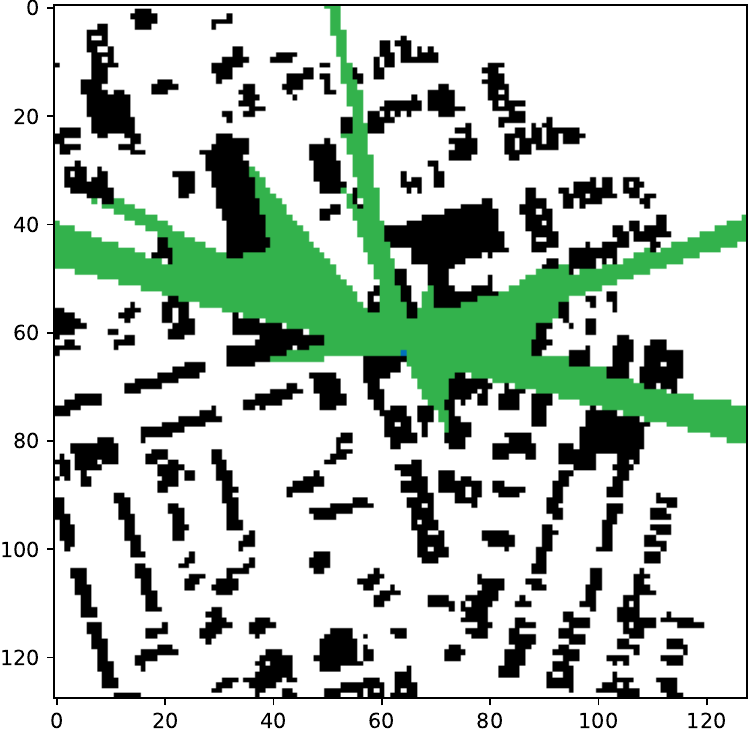}}\hfil
    \subfloat[\footnotesize{(b) NLoS (red)}]
    {\includegraphics[width=0.45\textwidth]{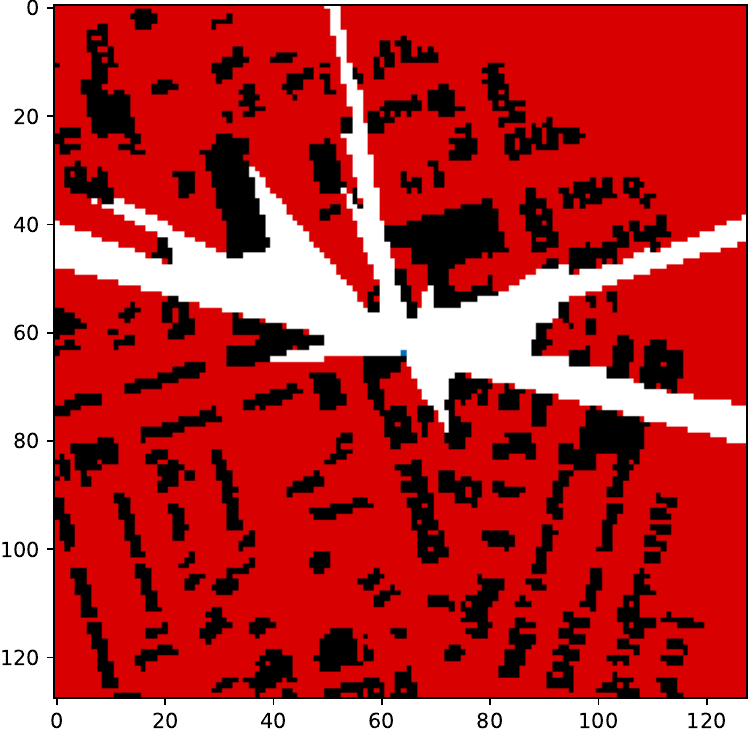}}\\
    \caption{(a) LoS and (b) NLoS masks corresponding to the building map in Fig.~\ref{fig:building_coverage}(a).}
    \label{fig:nloS_mask}
\end{minipage}
}
\end{figure}

The NLoS mask $\mathbf{M}_\text{NLoS}$ is defined as the complement of $\mathbf{M}_\text{LoS}$, with building interiors excluded. Hence, $\mathbf{M}_\text{LoS}$ identifies pixels with unobstructed paths to the Tx, while $\mathbf{M}_\text{NLoS}$ highlights free-space regions shadowed by buildings. Fig.~\ref{fig:nloS_mask} shows an example of the generated masks based on the building map in Fig.~\ref{fig:building_coverage}(a), where green pixels indicate LoS and red pixels indicate NLoS. During training, $\mathbf{M}_\text{NLoS}$ is incorporated as an additional input channel to improve feature learning in blocked areas.

\subsubsection{Sampled 3.5\,GHz Maps}\label{sec:3.5GHz_Sampled_Maps}
To simulate scenarios in which the complete 3.5\,GHz coverage map $\mathbf{S}_c$ is unavailable, three variants $\widetilde{\mathbf{S}}_c$ were generated to represent different sampling conditions:

\begin{itemize}
    \item \textbf{3.5\,GHz Sparse Map:} Two sampling strategies were adopted to construct sparse low-frequency maps with $N_c$ available pixels.
    \begin{itemize}
        \item \textit{Random sampling:} $N_c$ pixels were uniformly selected at random, and the remaining pixels were assigned $-160$\,dBm.

        \item \textit{NLoS-guided sampling:} The same number of pixels was selected, with a fraction $\gamma$ drawn from NLoS regions and $(1-\gamma)$ from LoS regions to emphasize blocked areas.
    \end{itemize}

    \item \textbf{3.5\,GHz Blend Map:}
    Starting from $N_c$ sampled pixels ($\gamma$ from NLoS and $(1-\gamma)$ from LoS regions), each selected value was expanded to its eight neighboring pixels (forming a nine-patch copy), resulting in $9N_c$ pseudo-points. Overlapping values were resolved by prioritizing SS values ${\geq -90}$\,dBm or by averaging. Building pixels were set to $-160$\,dBm. This approach increases sampling density while preserving local spatial gradients.

    \item \textbf{3.5\,GHz Block Map:}
    The map was divided into $10 \times 10$ sliding blocks (14,161 in total), and the 10 blocks with the highest NLoS ratios were selected. Some pixels within these blocks corresponded to building interiors, which were not regions of interest. To compensate for these non-usable pixels, an equal number of non-building pixels was additionally sampled outside the selected blocks. Each map thus contained $N_c$ sampled pixels, with greater emphasis on building edges and blockage regions.

\end{itemize}

\begin{figure}[!t]
\centering
{
\begin{minipage}{0.48\textwidth}
    \centering
    \captionsetup[subfloat]{labelformat=empty}
    \subfloat[\footnotesize{(a) 3.5\,GHz Blend Map (dBm)}]
    {\includegraphics[width=0.45\textwidth]{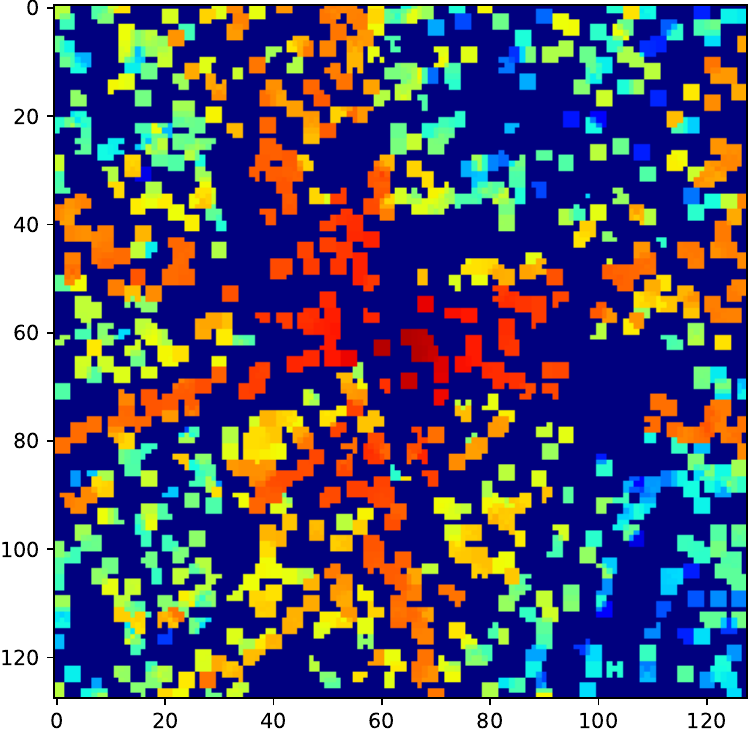}}\hfil
    \subfloat[\footnotesize{(b) 3.5\,GHz Block Map (dBm)}]
    {\includegraphics[width=0.45\textwidth]{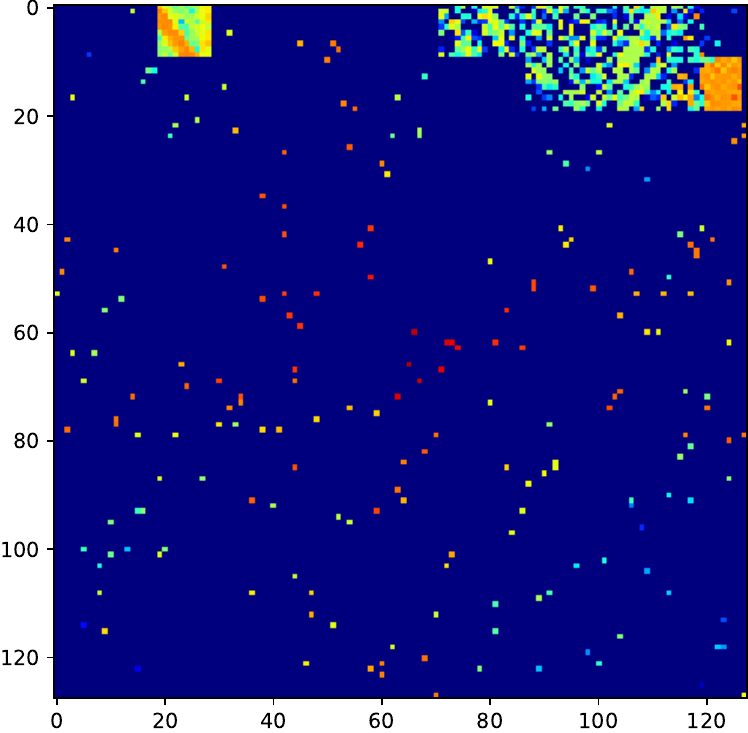}}\\
    \vspace{0.25cm}
    \includegraphics[width=0.8\textwidth]{Figs_pdf/fig4_colorbar}\\
    \vspace{-0.25cm}
    \caption{Examples of 3.5\,GHz (a) blend and (b) block maps derived from the coverage map in Fig.~\ref{fig:building_coverage}(b).}
    \label{fig:blend_block}
\end{minipage}
}
\end{figure}

Fig.~\ref{fig:blend_block} shows examples of blend and block maps generated from the coverage map in Fig.~\ref{fig:building_coverage}(b). The blend map preserves local gradients by expanding sparse samples, whereas the block map emphasizes NLoS-dominant regions around building edges.

\begin{table}[t]
\centering
{
\caption{Notation Summary}
\label{tab:notation_sec4}
\renewcommand{\arraystretch}{1.15}
\setlength{\tabcolsep}{6pt}
\begin{tabular}{p{0.1\linewidth} p{0.8\linewidth}}
\hline
\textbf{Symbol} & \textbf{Meaning} \\
\hline
$\mathbf{B}$ & Building height map (environment geometry) \\
$\mathbf{S_c}$ & Complete 3.5~GHz coverage map (SS map) \\
$\widetilde{\mathbf{S}}_c$ & Sampled/incomplete 3.5~GHz map (available low-band observation) \\
$\widehat{\mathbf{S}}_c$ & Reconstructed 3.5~GHz coverage map (coverage decoder output) \\
$\mathbf{S}_{d i}$ & Ground-truth 7~GHz SS map for direction $i$, $i\in\{1,\ldots,8\}$ \\
$\widehat{\mathbf{S}}_{d i}$ & Predicted 7~GHz SS map for direction $i$ \\
$\widetilde{\mathbf{S}}_{di}$ & Sparse-sampled 7~GHz SS map for direction $i$ (input anchor) \\
$\mathbf{X}$ & Available input tensor \\
$\mathbf{Y}$ & Prediction target (ground truth) \\
$\widehat{\mathbf{Y}}$ & Model prediction \\
$\mathbf{X}_{\text{Full}}$ & Full CommUNext input \\
$\mathbf{X}_{\text{Partial}}$ & Partial CommUNext input \\
$\mathbf{M}_{\text{NLoS}}$ & NLoS mask \\
$\widehat{\mathbf{M}}_{\text{NLoS}}$ & Predicted NLoS mask (segmentation branch output) \\
$L_{\text{Full}}$ & Full CommUNext training loss \\
$L_{\text{Partial}}$ & Partial CommUNext training loss \\
$L_{\text{SS}}$ & SS regression loss \\
$L_{\text{Seg}}$ & Segmentation loss \\
$L_{\text{Cov}}$ & 3.5~GHz coverage map reconstruction loss \\
$\lambda_{\text{Seg}}$ & Segmentation loss weight, $\lambda_{\text{Seg}}=0.3$ \\
$\lambda_{\text{Cov}}$ & Coverage reconstruction loss weight, $\lambda_{\text{Cov}}=0.5$ \\
\hline
\end{tabular}
}
\end{table}

\section{Data-Driven Framework}
\label{sec:CommUNext}

The proposed framework addresses cross-band and multi-directional signal prediction. Two scenarios are considered depending on the availability of 3.5\,GHz coverage data. In the Full CommUNext scenario, the complete 3.5\,GHz map $\mathbf{S}_c$ is available, providing strong reference information for cross-band learning. In the Partial CommUNext scenario, only sampled maps $\widetilde{\mathbf{S}}_c$ are accessible, and an auxiliary reconstruction strategy is introduced to compensate for missing data. This section first formulates the theoretical principle underlying cross-band and multi-directional prediction, and then presents the two neural architectures designed to approximate the Bayes-optimal solution. All notations are summarized in Table~\ref{tab:notation_sec4}.

\subsection{Problem Formulation and Theoretical Principle}
\label{sec:theory}

The task is formulated as a conditional estimation problem.
Let
\begin{equation}
\mathbf{X} = \left\{\mathbf{B}, \mathbf{S}_c \ (\text{or}\ \widetilde{\mathbf{S}}_c), \widetilde{\mathbf{S}}_{d1}, \dots, \widetilde{\mathbf{S}}_{d8} \right\},
\end{equation}
denote the available input tensor, where $\mathbf{B}$ is the building height map, $\mathbf{S}_c$ is the complete 3.5\,GHz coverage map, and $\{\widetilde{\mathbf{S}}_{di}\}$ are sparse directional 7\,GHz SS maps.
The prediction target is the set of complete 7\,GHz SS maps
\begin{equation}
\mathbf{Y} = \left\{\mathbf{S}_{d1}, \dots, \mathbf{S}_{d8} \right\}.
\end{equation}
The objective is to learn a predictor $f:\mathbf{X} \mapsto \widehat{\mathbf{Y}}$ that minimizes the expected mean squared error (MSE):
\begin{equation} \label{eq:metric_def}
\min_{f} \ \mathbb{E} \left[ \lVert \mathbf{Y} - f(\mathbf{X}) \rVert_F^2\right],
\end{equation}
where $\lVert \cdot \rVert_F$ denotes the Frobenius norm.\footnote{
The Frobenius norm of an $N$-th order tensor $\mathbf{Y} \in \mathbb{R}^{I_1 \times I_2 \times \cdots \times I_N}$ is defined as the square root of the sum of the squares of all its elements.
}

\begin{theorem}[MSE-Optimal Predictor under Squared Frobenius Loss~\cite{Kevin2012ML}]
\label{thm:full}
Consider the conditional estimation problem stated above.
Assume that:
i) $\mathbf{Y}$ is square-integrable, i.e., $\mathbb{E}\!\left[\|\mathbf{Y}\|_F^2\right]<\infty$; and
ii) the predictor $f$ is a measurable mapping satisfying
$\mathbb{E}\!\left[\|\mathbf{Y}-f(\mathbf{X})\|_F^2\right]<\infty$.
Then any predictor that minimizes the expected MSE in \eqref{eq:metric_def}
is given almost surely by the conditional expectation
\begin{equation}
f^\star(\mathbf{X})=\mathbb{E}[\mathbf{Y}\mid \mathbf{X}].
\end{equation}
\end{theorem}

Theorem~\ref{thm:full} establishes that the optimal predictor under the MSE criterion corresponds to the conditional mean of the target given all available observations.
Consequently, any component of $\mathbf{X}$ that is statistically informative of $\mathbf{Y}$ contributes to reducing the posterior uncertainty.
In particular, the low-frequency coverage map $\mathbf{S}_c$  provides cross-band structural information related to large-scale propagation geometry, while the sparse directional high-frequency maps $\{\widetilde{\mathbf{S}}_{di}\}$ offer complementary orientation-dependent observations that constrain the conditional expectation of each target map.

\begin{corollary}[Incomplete Reference-Band Case]
\label{cor:partial}
Suppose that the complete reference-band map $\mathbf{S}_c$ is unavailable and only its sparse observation $\widetilde{\mathbf{S}}_c$ is given.
The Bayes-optimal predictor depends on the latent $\mathbf{S}_c$ through its posterior distribution.
Consequently, the conditional mean
\begin{equation}
\widehat{\mathbf{S}}_c \triangleq \mathbb{E}\!\left[\mathbf{S}_c \mid \mathbf{B},\widetilde{\mathbf{S}}_c \right]
\end{equation}
is the MSE-optimal reconstruction of the missing reference.
This result provides the theoretical motivation for a two-stage estimation framework, where $\widehat{\mathbf{S}}_c$ is explicitly reconstructed to serve as a structural proxy for the latent reference information before performing the final target prediction.
\end{corollary}

\noindent\textbf{Proof:}
When only $\widetilde{\mathbf{S}}_c$ is available, the Bayes-optimal predictor in Theorem~\ref{thm:full} reduces to
\begin{equation}
f^\star(\mathbf{B}, \widetilde{\mathbf{S}}_c, \{\widetilde{\mathbf{S}}_{di}\})
=
\mathbb{E}\!\left[\mathbf{Y}\mid \mathbf{B}, \widetilde{\mathbf{S}}_c, \{\widetilde{\mathbf{S}}_{di}\}\right].
\end{equation}
Applying the law of iterated expectations with respect to the latent variable $\mathbf{S}_c$ yields
\begin{equation}
\label{eq:iterated}
\mathbb{E}[\mathbf{Y}\mid \mathbf{B},\widetilde{\mathbf{S}}_c,\{\widetilde{\mathbf{S}}_{di}\}]
=
\mathbb{E}\!\left[
\mathbb{E}[\mathbf{Y}\mid \mathbf{B},\mathbf{S}_c,\{\widetilde{\mathbf{S}}_{di}\}]
\;\Big|\;
\mathbf{B},\widetilde{\mathbf{S}}_c,\{\widetilde{\mathbf{S}}_{di}\}
\right].
\end{equation}
Let $g(\mathbf{B},\mathbf{S}_c,\{\widetilde{\mathbf{S}}_{di}\}) \triangleq \mathbb{E}[\mathbf{Y}\mid \mathbf{B},\mathbf{S}_c,\{\widetilde{\mathbf{S}}_{di}\}]$ denote the oracle predictor given full information, where we utilize the property that $\mathbf{S}_c$ subsumes the information in $\widetilde{\mathbf{S}}_c$.
Equation~\eqref{eq:iterated} indicates that the optimal solution averages the oracle prediction over the posterior distribution of the missing data.

In the proposed framework, we adopt a plug-in estimator strategy by substituting the latent $\mathbf{S}_c$ with its optimal estimate $\widehat{\mathbf{S}}_c$.
While exact equality $f^\star \approx g(\dots, \widehat{\mathbf{S}}_c, \dots)$ strictly holds only under specific conditions (e.g., if $g$ is affine in $\mathbf{S}_c$), this approach effectively approximates the integration over the posterior by injecting the most probable structural features.
Thus, recovering $\widehat{\mathbf{S}}_c$ acts as a necessary pretext task to constrain the hypothesis space for the downstream predictor $g$.
\hfill$\blacksquare$

\subsection{Full CommUNext}
\label{sec:FullCommUNext}

\begin{figure*} 
\centering
\includegraphics[width=\textwidth]{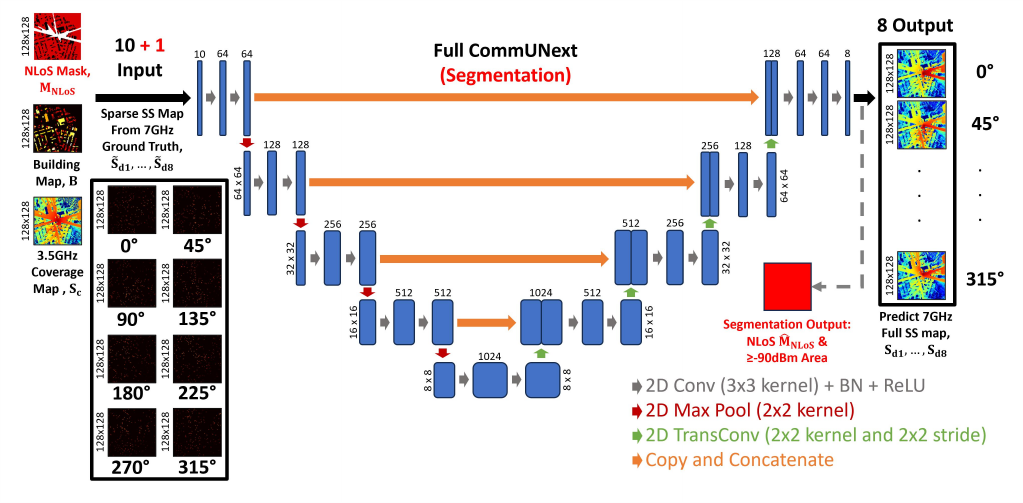}
\caption{Full CommUNext architecture.}
\label{fig:full_communext_seg}
\end{figure*}

As stated in Theorem~\ref{thm:full}, when the complete 3.5\,GHz coverage map $\mathbf{S}_c$ is available, the optimal predictor is
${f^\star(\mathbf{B},\mathbf{S}_c,\{\widetilde{\mathbf{S}}_{di}\}) = \mathbb{E}{[\mathbf{Y} \mid \mathbf{B},\mathbf{S}_c,\{\widetilde{\mathbf{S}}_{di}\}]}}$.
The Full CommUNext network is designed to approximate this conditional expectation using a U-Net backbone extended for cross-band and multi-directional prediction.

The architecture of Full CommUNext is illustrated in Fig. \ref{fig:full_communext_seg}.
The input tensor is defined as
\begin{equation}
\mathbf{X}_\text{Full} = \left\{\mathbf{B}, \mathbf{S}_c, \widetilde{\mathbf{S}}_{d1}, \dots, \widetilde{\mathbf{S}}_{d8} \right\}.
\end{equation}
Full CommUNext adopts an encoder-decoder backbone. The encoder applies stacked $2$D convolutions with batch normalization and ReLU activations, interleaved with max-pooling, to extract hierarchical multi-scale features that jointly encode environmental geometry and propagation characteristics. The decoder mirrors this process using transposed convolutions and skip connections to fuse low-level spatial details with high-level semantic representations, enabling accurate reconstruction of fine-grained SS patterns while preserving global consistency. The final output of the network is a set of eight reconstructed high-band SS maps,
\begin{equation}
\widehat{\mathbf{Y}} = \{\widehat{\mathbf{S}}_{d1}, \dots, \widehat{\mathbf{S}}_{d8}\},
\end{equation}
corresponding to the full azimuthal coverage.

To better handle blocked regions, Full CommUNext incorporates a segmentation branch that integrates a structured prior based on environmental geometry. The NLoS mask $\mathbf{M}_{\mathrm{NLoS}}$ is appended as an additional input channel, while an auxiliary segmentation branch predicts $\widehat{\mathbf{M}}_{\mathrm{NLoS}}$ to identify communication-available NLoS regions. By jointly learning SS reconstruction and NLoS segmentation, the model constrains the feasible support of $\mathbf{Y}$, reducing prediction uncertainty and improving robustness in shadowed areas.

The training, validation, and testing procedures are summarized below, following the dataset partitioning in Section~\ref{sec:data_partitioning}.

\subsubsection{Training}
The model is trained for $500$ epochs using the Adam optimizer with an initial learning rate of $10^{-3}$ and a batch size of $32$. Each epoch corresponds to one complete pass over the training dataset, where each sample is an $(\mathbf{X}, \mathbf{Y})$ pair, and the data are randomly shuffled prior to batching. The learning rate is halved if the validation loss does not improve for $5$ consecutive epochs.
The overall training objective combines regression and segmentation terms:
\begin{equation} \label{eq:L_full}
L_\text{Full}
= L_\text{SS} + \lambda_\text{Seg}\,L_\text{Seg},
\end{equation}
where $L_\text{SS}$ represents the MSE between the predicted high-band SS maps ${\widehat{\mathbf{Y}} = \{ \widehat{\mathbf{S}}_{di} \}}$ and ground-truth SS maps ${\mathbf{Y} = \{ \mathbf{S}_{di} \}}$,
and $L_\text{Seg}$ denotes the binary cross-entropy (BCE) between the predicted and ground-truth NLoS masks. The weighting factor $\lambda_\text{Seg}=0.3$ is a hyperparameter that controls the contribution of the segmentation task.

Notably, $L_\text{SS}$ is computed in the logarithmic (dB) domain rather than the linear power domain. This choice accounts for the large dynamic range of wireless SS, as linear-domain losses are dominated by strong-signal regions and tend to undervalue errors in weak-signal areas. The dB-domain formulation treats relative errors more uniformly and leads to more balanced learning across coverage-limited regions.

\subsubsection{Validation}
A held-out validation set is used for model selection. The same input configuration and sampling strategy as in training are applied, and the checkpoint with the best validation performance is selected.

\subsubsection{Testing}
The selected checkpoint is evaluated on unseen scenes for testing. All methods are compared under identical test sets and sampling strategies. Detailed results are reported in Section~\ref{sec:experimental_result}.

In summary, Full CommUNext realizes the Bayes predictor of Theorem~\ref{thm:full} in practice by fusing complete low-frequency priors with sparse high-frequency directional anchors, while the \textbf{Seg} variant incorporates $\mathbf{M}_\text{NLoS}$ as a structured prior to enhance robustness in NLoS regions.

\begin{remark}[Structured Prior via NLoS Mask]
The NLoS mask $\mathbf{M}_\text{NLoS}$ provides structured prior knowledge by explicitly identifying shadowed but communication-available regions.
Formally, conditioning on $\mathbf{M}_\text{NLoS}$ reduces the predictive uncertainty of $Y$:
\begin{equation}
\mathrm{Var}(\mathbf{Y}\mid \mathbf{X},\mathbf{M}_\text{NLoS})
\;\leq\; \mathrm{Var}(\mathbf{Y}\mid \mathbf{X}),
\end{equation}
where $\mathbf{X}$ denotes the available inputs (either $\mathbf{X}_\text{Full}$ or $\mathbf{X}_\text{Partial}$).
The inequality follows from the law of total variance, since $\mathbf{M}_\text{NLoS}$ refines the conditional distribution of $\mathbf{Y}$.
In practice, incorporating $\mathbf{M}_\text{NLoS}$ as an input channel and training an auxiliary segmentation branch constrain predictions to feasible communication regions, improving robustness and reducing error in blocked areas.
\end{remark}

\subsection{Partial CommUNext}
\label{sec:PartialCommUNext}

\begin{figure*} 
\centering
\includegraphics[width=\textwidth]{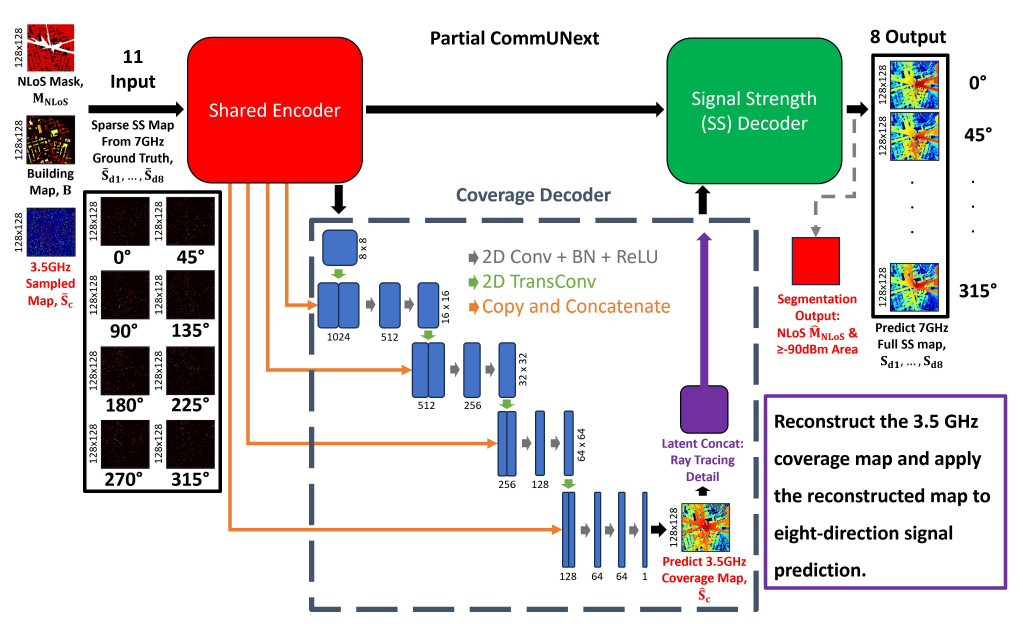}
\caption{Partial CommUNext architecture.}
\label{fig:partial_communext}
\end{figure*}

Corollary~\ref{cor:partial} shows that when $\mathbf{S}_c$ is unavailable, the optimal predictor depends on $\{\mathbf{B}, \widetilde{\mathbf{S}}_c, \widetilde{\mathbf{S}}_{di}\}$, with latent reconstruction of $\mathbf{S}_c$ required through ${\widehat{\mathbf{S}}_c = \mathbb{E}[\mathbf{S}_c\mid \mathbf{B},\widetilde{\mathbf{S}}_c]}$. A straightforward approach would be to design a separate coverage-reconstruction network and then feed the reconstructed 3.5\,GHz map into the Full CommUNext model.
However, this two-stage process is inefficient because the latent features used to reconstruct $\mathbf{S}_c$ are not directly leveraged for high-band prediction. Inspired by the dual-decoder design of Y-Net~\cite{mehta2018ynet}, the Partial CommUNext architecture introduces a coverage-reconstruction branch alongside the main SS prediction branch, as illustrated in Fig.~\ref{fig:partial_communext}.

The input tensor is
\begin{equation}
\mathbf{X}_\text{Partial} = \{\mathbf{B}, \widetilde{\mathbf{S}}_c, \widetilde{\mathbf{S}}_{d1}, \dots, \widetilde{\mathbf{S}}_{d8}\},
\end{equation}
where $\widetilde{\mathbf{S}}_c$ is a sampled 3.5\,GHz map (Sec.~\ref{sec:3.5GHz_Sampled_Maps}).
The encoder is identical to Full CommUNext, ensuring consistent feature extraction. At the bottleneck, two decoders operate jointly:
\begin{itemize}
    \item \textbf{Coverage Decoder:} Reconstructs $\widehat{\mathbf{S}}_c$ from $\widetilde{\mathbf{S}}_c$, serving as an auxiliary task that approximates $\mathbb{E}[\mathbf{S}_c\mid \mathbf{B},\widetilde{\mathbf{S}}_c]$ and transfers latent propagation knowledge.
    \item \textbf{SS Decoder:} Predicts the complete 7\,GHz outputs $\widehat{\mathbf{Y}}$,
    using skip connections and the reconstructed $\widehat{\mathbf{S}}_c$ as an auxiliary input.
    It also includes a segmentation branch that outputs $\widehat{\mathbf{M}}_\text{NLoS}$, consistent with Full CommUNext.
\end{itemize}

Following the same training configuration described in Section~\ref{sec:FullCommUNext}, the Partial CommUNext model is optimized using a single-stage, end-to-end objective that extends the Full CommUNext loss by incorporating an additional coverage reconstruction term:
\begin{equation} \label{eq:L_partial}
L_\text{Partial}
= L_\text{SS}
+ \lambda_\text{Seg}\,L_\text{Seg}
+ \lambda_\text{Cov}\,L_\text{Cov},
\end{equation}
where $L_\text{Cov}$ is the MSE between $\widehat{\mathbf{S}}_c$ and $\mathbf{S}_c$. The weighting factor $\lambda_\text{Cov}=0.5$ is treated as a hyperparameter controlling the relative contribution of the coverage reconstruction objective and is selected based on validation performance.
All loss terms are jointly back-propagated in each training iteration to update the model parameters.

In summary, Partial CommUNext operationalizes Corollary~\ref{cor:partial} by combining sampled low-band inputs with auxiliary coverage reconstruction, enabling effective approximation of the latent $\mathbf{S}_c$ and thereby improving high-band SS prediction even under incomplete reference conditions.

\section{Experimental Results} \label{sec:experimental_result}
This section presents the analysis and discussion of the prediction performance of the proposed Full CommUNext and Partial CommUNext architectures.
Evaluation is conducted by comparing the model predictions on the test set with the corresponding ground truth.
Two error metrics are adopted: the mean absolute error (MAE) and the root mean squared error (RMSE):
\begin{subequations}
    \begin{align}
        \mathrm{MAE} &= \frac{1}{8 N}\sum_{i=1}^{8} \sum_{n=1}^{N} \left|\widehat{S}_{di,n} - S_{di,n}  \right|,~~(\text{dB})
        \label{eq:mae} \\
        \mathrm{RMSE} &= \sqrt{ \frac{1}{8 N}\sum_{i=1}^{8} \sum_{n=1}^{N} \left|\widehat{S}_{di,n} - S_{di,n}  \right|^2 },~~(\text{dB})  \label{eq:rmse}
    \end{align}
\end{subequations}
where $\widehat{S}_{di,n}$ and $S_{di,n}$ denote the predicted and ground-truth dBm values of the $n$-th pixel in $\mathbf{S}_{di}$, respectively, and $N$ represents the number of pixels of each ground-truth map.

\begin{figure} 
\centering
\includegraphics[width=0.45\textwidth]{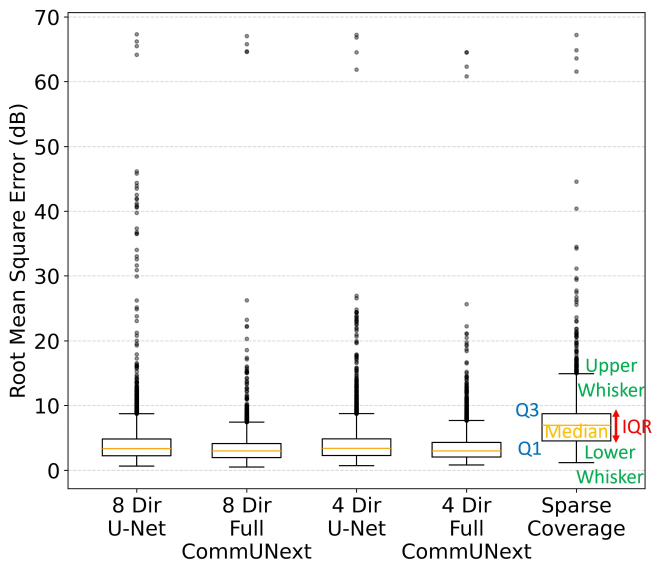}
\caption{Box plots of RMSE for Full CommUNext under different numbers of directional inputs.}
\label{fig:ModOptResults}
\end{figure}

Beyond average performance, the statistical distribution of prediction errors is examined using box plots of MAE and RMSE (see Fig.~\ref{fig:ModOptResults} as an example).
As illustrated in Fig.~\ref{fig:ModOptResults}, the interquartile range (IQR) quantifies the variability of errors and is defined as the difference between the third quartile ($Q_3$) and the first quartile ($Q_1$):
\begin{equation}
\mathrm{IQR} = Q_{3} - Q_{1},
\end{equation}
representing the range covering the central 50\% of the data.
A larger IQR indicates greater dispersion of the main error distribution, whereas a smaller IQR implies that errors are more concentrated around the median.
In box plots, the lower whisker (L-Whisker) extends to the smallest point within ${Q_1 - 1.5\times\mathrm{IQR}}$, and the upper whisker (U-Whisker) extends to the largest point within ${Q_3 + 1.5\times\mathrm{IQR}}$.
Data points outside this range are considered outliers and typically correspond to rare but extreme prediction errors.\footnote{Only pixels within the communicable region are analyzed in the box plots, since including outliers in non-communicable regions is meaningless.
A communicable point is defined as a pixel satisfying ${S_{di,n} \geq -90}$\,dBm, corresponding to the approximate thermal noise level at 300~K, 100~MHz bandwidth, and 4~dB noise figure.}

From a practical standpoint, a large difference between the IQRs of MAE and RMSE indicates that outliers significantly skew the error distribution.
Conversely, similar IQRs imply a more uniform error distribution and better model robustness to extreme deviations.
Therefore, joint consideration of MAE, RMSE, and IQR provides a comprehensive assessment of both the accuracy and stability of the proposed architectures.

In the Full CommUNext experiments, the model’s prediction capability is evaluated using the complete 3.5\,GHz coverage map as input.
Three aspects are analyzed in detail: (i) performance before and after architectural optimization, (ii) the effect of reducing the number of 7\,GHz directional inputs, and (iii) the impact of different 7\,GHz sampling strategies.
For the Partial CommUNext experiments, the focus is on assessing the reconstruction of the 3.5\,GHz coverage map under incomplete coverage conditions.
By generating input maps through various sparse-sampling strategies, we further analyze how different sampling patterns affect the overall prediction accuracy and robustness of the model.

\subsection{Full CommUNext (With Complete 3.5\,GHz Coverage Map)}
\subsubsection{Model Optimization Comparison}
\label{sec:ModelOpt}
Full CommUNext is evaluated using the network introduced in Section~\ref{sec:FullCommUNext}.
The baseline U-Net employs only the regression loss $L_\text{SS}$ in \eqref{eq:L_full}, focusing solely on the prediction accuracy of SS maps.
The extended Full CommUNext architecture, by contrast, incorporates the NLoS mask $\mathbf{M}_\text{NLoS}$ as an additional input channel and activates the segmentation branch.
In this configuration, $L_\text{Seg}$ is included with a weighting factor of $\lambda_\text{Seg}=0.3$, resulting in a multi-task objective that jointly optimizes SS reconstruction and NLoS mask prediction.

{
Six experimental settings are designed to evaluate the effectiveness of the Full CommUNext framework and to examine its core hypothesis, namely whether reliable high-frequency SS prediction can still be achieved when a subset of directional inputs is unavailable by leveraging low-frequency information as guidance.
\begin{enumerate}
\item[i.] \textbf{Kriging:} The Kriging algorithm~\cite{Koya2017Kriging} is adopted as a classical interpolation-based baseline.

\item[ii.] \textbf{8 Dir U-Net:} All eight 7\,GHz sparse SS maps are provided as inputs to a vanilla U-Net without the segmentation branch.

\item[iii.] \textbf{8 Dir Full CommUNext:} Same as (ii), but with the segmentation branch enabled.

\item[iv.] \textbf{4 Dir U-Net:} Only four sparse SS maps ($0^\circ$, $90^\circ$, $180^\circ$, and $270^\circ$) are used as directional inputs.

\item[v.] \textbf{4 Dir Full CommUNext:} Same as (iv), but with the segmentation branch enabled.

\item[vi.] \textbf{Sparse Coverage:} The complete 3.5\,GHz coverage map is unavailable; only sparsely sampled 3.5\,GHz coverage information and four 7 GHz sparse SS maps is provided as input.
\end{enumerate}
In these experiments, each 7\,GHz sparse SS map $\widetilde{\mathbf{S}}_{di}$ contains $N_d = 200$ randomly sampled pixels. When only sparsely sampled 3.5\,GHz coverage information is available, each sparse 3.5\,GHz SS map $\widetilde{\mathbf{S}}_c$ contains $N_c = 1{,}000$ randomly sampled pixels.}

\begin{table*}
\centering
\scriptsize
\captionof{table}{Model optimization results within the communicable region: MAE and RMSE.}
\label{tab:mae_rmse_results}
\begin{tabular}{lcccccc|cccccc}
\hline
& \multicolumn{6}{c|}{\textbf{MAE}} & \multicolumn{6}{c}{\textbf{RMSE}} \\
\textbf{Case} & \textbf{Q1} & \textbf{Q3} & \textbf{IQR} & \textbf{Median} & \textbf{L-Whisker} & \textbf{U-Whisker} & \textbf{Q1} & \textbf{Q3} & \textbf{IQR} & \textbf{Median} & \textbf{L-Whisker} & \textbf{U-Whisker} \\
\hline
Kriging
& 2.460
& 8.537
& 6.077
& 5.773
& 0.00
& 17.243
& 4.965
& 11.601
& 6.636
& 8.679
& 0.000
& 21.424 \\
8 Dir U-Net           & 1.227 & 2.641 & 1.414 & 1.797 & 0.461 & 4.759 & 2.264 & 4.854 & 2.589 & 3.367 & 0.657 & 8.727 \\
\rowcolor{gray!15}
8 Dir Full CommUNext      & 1.037 & 2.076 & 1.039 & 1.489 & 0.409 & 3.634 & 1.974 & 4.161 & 2.187 & 3.011 & 0.518 & 7.426 \\
4 Dir U-Net           & 1.297 & 2.679 & 1.383 & 1.859 & 0.520 & 4.752 & 2.292 & 4.873 & 2.581 & 3.411 & 0.722 & 8.744 \\
4 Dir Full CommUNext      & 1.305 & 2.421 & 1.116 & 1.717 & 0.627 & 4.094 & 2.076 & 4.316 & 2.240 & 3.011 & 0.844 & 7.672 \\
Sparse Coverage  & 2.563 & 5.631 & 3.069 & 4.371 & 0.924 & 10.202 & 4.550 & 8.728 & 4.178 & 6.940 & 1.205 & 14.907 \\
\hline
\end{tabular}
\end{table*}

\begin{figure*}[!t]
\centering
\captionsetup[subfloat]{labelformat=empty}
\subfloat[\footnotesize{(a) Building 1: \\Ground Truth}]
{\includegraphics[width=0.13\textwidth]{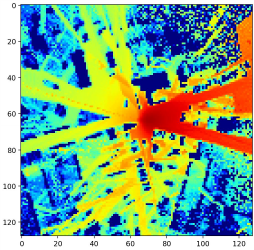}}\hfil
\subfloat[\footnotesize{(b) 4 Dir Full CommUNext\\ RMSE = 7.365\,dB\\ MAE = 3.339\,dB}]
{\includegraphics[width=0.13\textwidth]{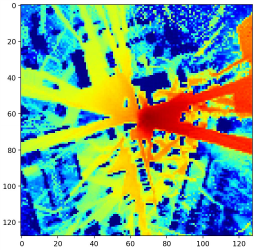}}\hfil
\subfloat[\footnotesize{(c) Sparse Coverage\\ RMSE = 12.243\,dB\\ MAE = 6.953\,dB}]
{\includegraphics[width=0.13\textwidth]{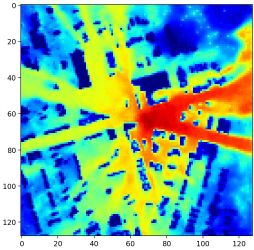}}\hfil
\subfloat[\footnotesize{(d) 0 Dir Full CommUNext\\ RMSE = 7.508\,dB\\ MAE = 3.467\,dB}]
{\includegraphics[width=0.13\textwidth]{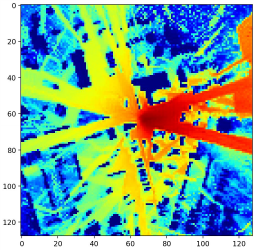}}\hfil
\subfloat[\footnotesize{(e) 1000\\ RMSE = 7.358\,dB\\ MAE = 2.717\,dB}]
{\includegraphics[width=0.13\textwidth]{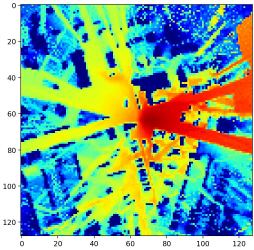}}\hfil
\subfloat[\footnotesize{(f) Block Coverage\\ RMSE = 9.061\,dB\\ MAE = 4.522\,dB}]
{\includegraphics[width=0.13\textwidth]{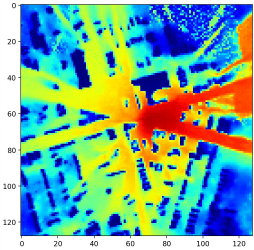}}

\subfloat[\footnotesize{(g) Building 2: \\Ground Truth}]
{\includegraphics[width=0.13\textwidth]{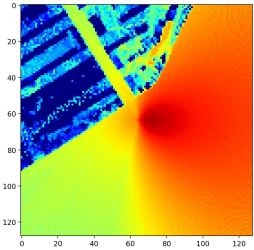}}\hfil
\subfloat[\footnotesize{(h) 4 Dir Full CommUNext\\ RMSE = 1.357\,dB\\ MAE = 1.133\,dB}]
{\includegraphics[width=0.13\textwidth]{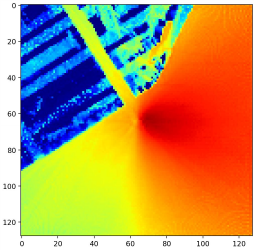}}\hfil
\subfloat[\footnotesize{(i) Sparse Coverage\\ RMSE = 2.015\,dB\\ MAE = 1.392\,dB}]
{\includegraphics[width=0.13\textwidth]{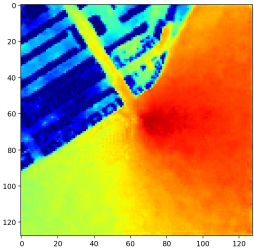}}\hfil
\subfloat[\footnotesize{(j) 0 Dir Full CommUNext\\ RMSE = 1.261\,dB\\ MAE = 0.975\,dB}]
{\includegraphics[width=0.13\textwidth]{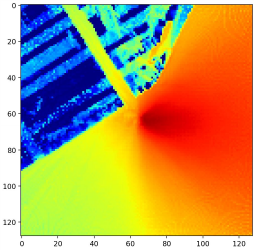}}\hfil
\subfloat[\footnotesize{(k) 1000\\ RMSE = 0.475\,dB\\ MAE = 0.301\,dB}]
{\includegraphics[width=0.13\textwidth]{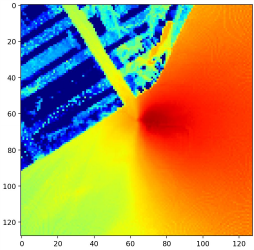}}\hfil
\subfloat[\footnotesize{(l) Block Coverage\\ RMSE = 1.287\,dB\\ MAE = 0.878\,dB}]
{\includegraphics[width=0.13\textwidth]{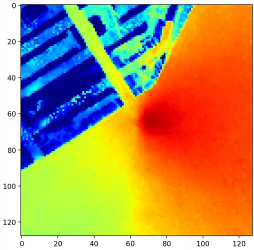}}

\subfloat[\footnotesize{(m) Building 3: \\Ground Truth}]
{\includegraphics[width=0.13\textwidth]{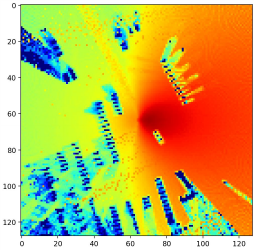}}\hfil
\subfloat[\footnotesize{(n) 4 Dir Full CommUNext\\ RMSE = 3.200\,dB\\ MAE = 1.391\,dB}]
{\includegraphics[width=0.13\textwidth]{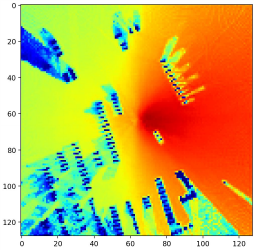}}\hfil
\subfloat[\footnotesize{(o) Sparse Coverage\\ RMSE = 5.483\,dB\\ MAE = 3.066\,dB}]
{\includegraphics[width=0.13\textwidth]{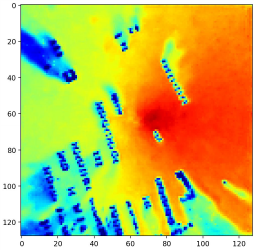}}\hfil
\subfloat[\footnotesize{(p) 0 Dir Full CommUNext\\ RMSE = 3.422\,dB\\ MAE = 1.421\,dB}]
{\includegraphics[width=0.13\textwidth]{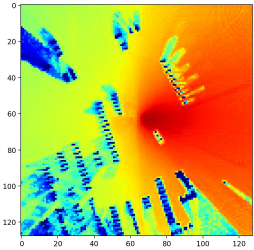}}\hfil
\subfloat[\footnotesize{(q) 1000\\ RMSE = 2.875\,dB\\ MAE = 0.881\,dB}]
{\includegraphics[width=0.13\textwidth]{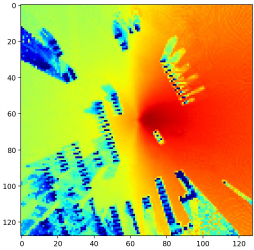}}\hfil
\subfloat[\footnotesize{(r) Block Coverage\\ RMSE = 4.360\,dB\\ MAE = 1.721\,dB}]
{\includegraphics[width=0.13\textwidth]{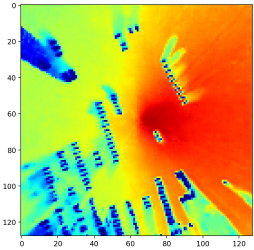}}

\includegraphics[width=0.9\textwidth]{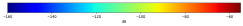}

\vspace{-0.35cm}

\caption{Comparison between ground truth and predictions under different conditions for three test maps (restricted to ground-truth regions with SS $\geq -90$\,dBm).}
\label{fig:test_samples}
\end{figure*}

\begin{figure*}[!t]
\centering
\captionsetup[subfloat]{labelformat=empty}

\subfloat[\footnotesize{(a) 4 Dir Full CommUNext\\ RMSE = 7.365\,dB\\ MAE = 3.339\,dB}]
{\includegraphics[width=0.13\textwidth]{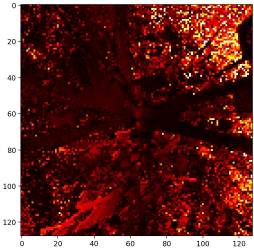}}\hfil
\subfloat[\footnotesize{(b) Sparse Coverage\\ RMSE = 12.243\,dB\\ MAE = 6.953\,dB}]
{\includegraphics[width=0.13\textwidth]{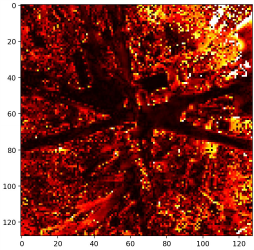}}\hfil
\subfloat[\footnotesize{(c) 0 Dir Full CommUNext\\ RMSE = 7.508\,dB\\ MAE = 3.467\,dB}]
{\includegraphics[width=0.13\textwidth]{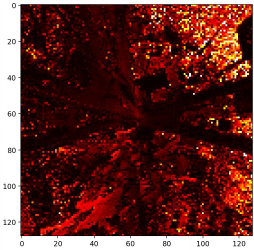}}\hfil
\subfloat[\footnotesize{(d) 1000\\ RMSE = 7.358\,dB\\ MAE = 2.717\,dB}]
{\includegraphics[width=0.13\textwidth]{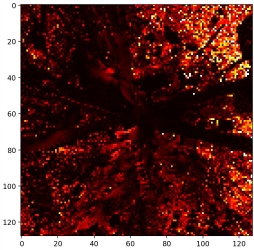}}\hfil
\subfloat[\footnotesize{(e) Block Coverage\\ RMSE = 9.061\,dB\\ MAE = 4.522\,dB}]
{\includegraphics[width=0.13\textwidth]{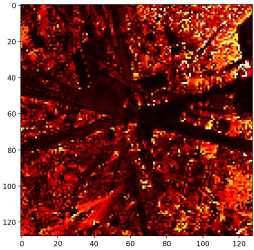}}

\subfloat[\footnotesize{(f) 4 Dir Full CommUNext\\ RMSE = 1.357\,dB\\ MAE = 1.133\,dB}]
{\includegraphics[width=0.13\textwidth]{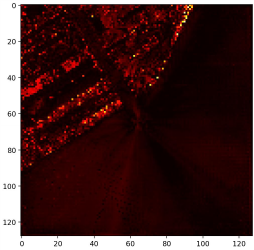}}\hfil
\subfloat[\footnotesize{(g) Sparse Coverage\\ RMSE = 2.015\,dB\\ MAE = 1.392\,dB}]
{\includegraphics[width=0.13\textwidth]{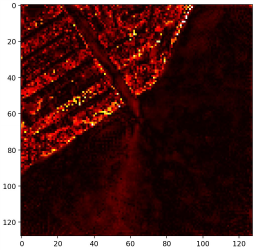}}\hfil
\subfloat[\footnotesize{(h) 0 Dir Full CommUNext\\ RMSE = 1.261\,dB\\ MAE = 0.975\,dB}]
{\includegraphics[width=0.13\textwidth]{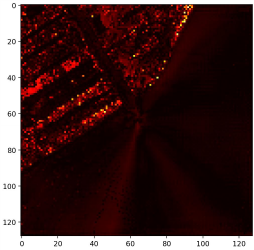}}\hfil
\subfloat[\footnotesize{(i) 1000\\ RMSE = 0.475\,dB\\ MAE = 0.301\,dB}]
{\includegraphics[width=0.13\textwidth]{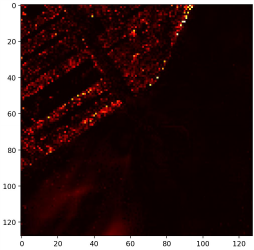}}\hfil
\subfloat[\footnotesize{(j) Block Coverage\\ RMSE = 1.287\,dB\\ MAE = 0.878\,dB}]
{\includegraphics[width=0.13\textwidth]{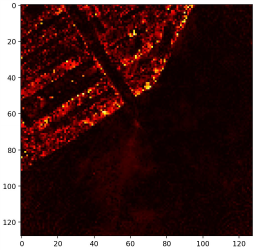}}

\subfloat[\footnotesize{(k) 4 Dir Full CommUNext\\ RMSE = 3.200\,dB\\ MAE = 1.391\,dB}]
{\includegraphics[width=0.13\textwidth]{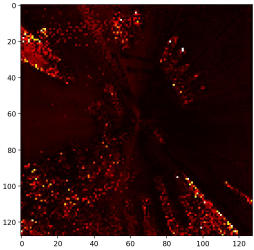}}\hfil
\subfloat[\footnotesize{(l) Sparse Coverage\\ RMSE = 5.483\,dB\\ MAE = 3.066\,dB}]
{\includegraphics[width=0.13\textwidth]{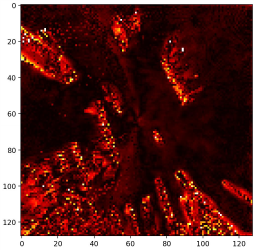}}\hfil
\subfloat[\footnotesize{(m) 0 Dir Full CommUNext\\ RMSE = 3.422\,dB\\ MAE = 1.421\,dB}]
{\includegraphics[width=0.13\textwidth]{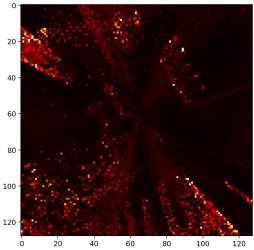}}\hfil
\subfloat[\footnotesize{(n) 1000\\ RMSE = 2.875\,dB\\ MAE = 0.881\,dB}]
{\includegraphics[width=0.13\textwidth]{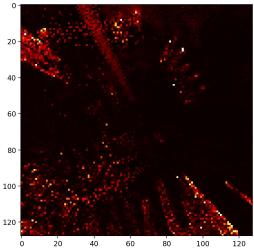}}\hfil
\subfloat[\footnotesize{(o) Block Coverage\\ RMSE = 4.360\,dB\\ MAE = 1.721\,dB}]
{\includegraphics[width=0.13\textwidth]{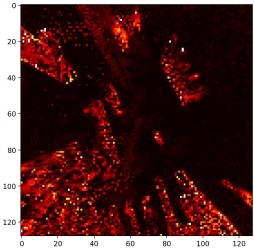}}

\includegraphics[width=0.9\textwidth]{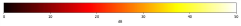}
\vspace{-0.35cm}
\caption{Absolute error maps under different conditions for three test maps. The absolute error map is defined pixel-wise as $e_{di,n} = | \widehat{S}_{di,n} - S_{di,n} |$ (dB), where $\widehat{S}_{di,n}$ and $S_{di,n}$ denote the predicted and ground-truth SS values at pixel $n$ in the $i$-th directional SS map, respectively.
}
\label{fig:error_maps}
\end{figure*}

To illustrate the practical behavior of the proposed framework, three representative cases are selected from 1,479 test maps, covering dense urban, single-block, and sparse-building scenarios. Fig.~\ref{fig:test_samples} compares the predicted 7\,GHz SS maps at $0^\circ$ with the corresponding ground truth, while Fig.~\ref{fig:error_maps} shows the associated absolute error maps defined as
\begin{equation}
e_{di,n} = \left| \widehat{S}_{di,n} - S_{di,n} \right| \quad (\mathrm{dB}),
\label{eq:error_map}
\end{equation}
where $e_{di,n}$ denotes the pixelwise absolute error. Darker colors indicate smaller errors, whereas lighter colors from red to yellow to white represent larger errors. Overall, the predictions are accurate in most areas, particularly along LoS paths and antenna beam lobes. Most large errors occur in NLoS open areas and near building edges within NLoS regions.

\begin{figure} 
\centering
\captionsetup[subfloat]{labelformat=empty}
\subfloat[\footnotesize{(a) 4 Dir Full CommUNext}]
{\includegraphics[width=0.45\columnwidth]{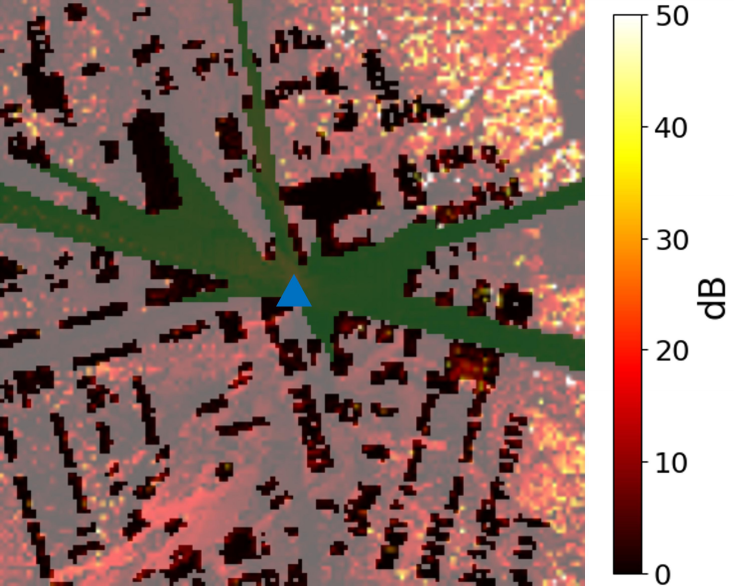}}\hfil
\subfloat[\footnotesize{(b) Sparse Coverage}]
{\includegraphics[width=0.45\columnwidth]{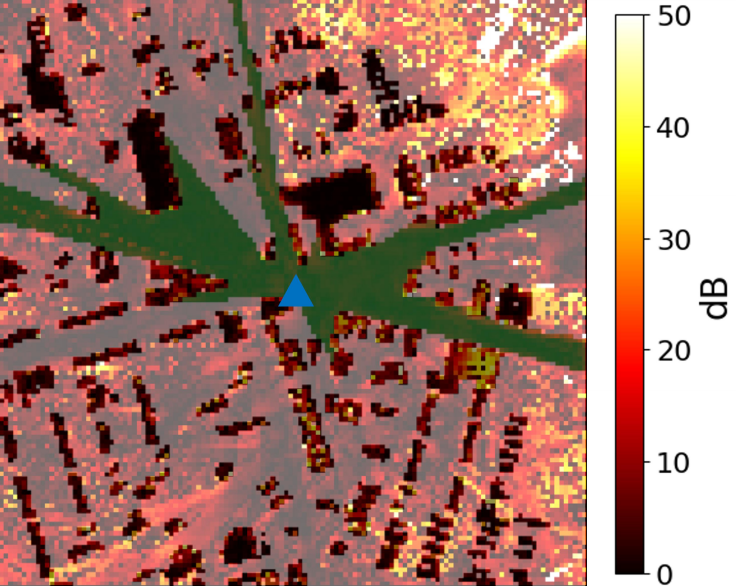}}
\caption{Spatial distribution of absolute prediction error for (a) \textbf{4 Dir Full CommUNext} and (b) \textbf{Sparse Coverage} under the same test scenario.
The green region indicates the LoS area, and the blue triangle marks the BS location.}
\label{fig:spatial_err}
\end{figure}

The \textbf{Sparse Coverage} results in Fig.~\ref{fig:test_samples}(c), (i), and (o) further demonstrate noticeable degradation when complete 3.5\,GHz coverage maps are unavailable, especially in NLoS open areas and at the boundaries between strong and weak beam directions. The corresponding error maps in Fig.~\ref{fig:error_maps}(b), (g), and (l) reveal elevated errors in these regions.
To further quantify the error distribution, Fig.~\ref{fig:error_maps}(a) and (b) are enlarged and the LoS regions are identified in Fig.~\ref{fig:spatial_err}(a) and (b). The MAE values in the LoS and NLoS regions are 1.55~dB and 3.57~dB for Fig.~\ref{fig:spatial_err}(a), and 2.28~dB and 8.39~dB for Fig.~\ref{fig:spatial_err}(b), respectively. These quantitative results confirm that prediction errors are predominantly concentrated in NLoS regions, highlighting the critical role of complete low-frequency coverage in constraining cross-band uncertainty.

\vspace{6pt}

\begin{table*}
\centering
\scriptsize
\captionof{table}{Directional (angle) comparison within the communicable region: MAE and RMSE.}
\label{tab:dir_mae_rmse}
\begin{tabular}{lcccccc|cccccc}
\hline
& \multicolumn{6}{c|}{\textbf{MAE}} & \multicolumn{6}{c}{\textbf{RMSE}} \\
\textbf{Case} & \textbf{Q1} & \textbf{Q3} & \textbf{IQR} & \textbf{Median} & \textbf{L-Whisker} & \textbf{U-Whisker} & \textbf{Q1} & \textbf{Q3} & \textbf{IQR} & \textbf{Median} & \textbf{L-Whisker} & \textbf{U-Whisker} \\
\hline
\rowcolor{gray!15}
4 Dir     & 1.305 & 2.421 & \bf{1.115} & 1.716 & 0.627 & 4.094 & 2.075 & 4.315 & 2.240 & 3.011 & 0.843 & 7.671 \\
2 Dir     & 1.176 & 2.424 & 1.248 & 1.674 & 0.428 & 4.296 & 2.068 & 4.507 & 2.438 & 3.131 & 0.605 & 8.164 \\
1 Dir     & 1.110 & 2.349 & 1.238 & 1.599 & 0.455 & 4.204 & 1.981 & 4.397 & 2.416 & 3.042 & 0.632 & 8.016 \\
1 Random  & 1.044 & 2.318 & 1.273 & 1.550 & 0.342 & 4.225 & 1.959 & 4.391 & 2.432 & 3.038 & 0.472 & 8.036 \\
0 Dir     & 1.265 & 2.483 & 1.218 & 1.744 & 0.566 & 4.309 & 2.150 & 4.577 & 2.426 & 3.182 & 0.753 & 8.201 \\
\hline
\end{tabular}
\end{table*}

\begin{table*}
\centering
\scriptsize
\captionof{table}{NLoS sparse sampling comparison within the communicable region: MAE and RMSE.}
\label{tab:nloS_mae_rmse}
\begin{tabular}{lcccccc|cccccc}
\hline
& \multicolumn{6}{c|}{\textbf{MAE}} & \multicolumn{6}{c}{\textbf{RMSE}} \\
\textbf{Case} & \textbf{Q1} & \textbf{Q3} & \textbf{IQR} & \textbf{Median} & \textbf{L-Whisker} & \textbf{U-Whisker} & \textbf{Q1} & \textbf{Q3} & \textbf{IQR} & \textbf{Median} & \textbf{L-Whisker} & \textbf{U-Whisker} \\
\hline
200  & 0.721 & 1.864 & 1.142 & 1.209 & 0.112 & 3.578 & 1.739 & 4.046 & 2.306 & 2.832 & 0.153 & 7.501 \\
400  & 0.721 & 1.805 & 1.084 & 1.190 & 0.158 & 3.431 & 1.702 & 3.911 & 2.208 & 2.751 & 0.215 & 7.219 \\
600  & 0.719 & 1.744 & 1.024 & 1.171 & 0.165 & 3.279 & 1.665 & 3.770 & 2.105 & 2.670 & 0.222 & 6.922 \\
800  & 0.719 & 1.729 & 1.009 & 1.169 & 0.174 & 3.241 & 1.645 & 3.695 & 2.049 & 2.623 & 0.239 & 6.760 \\
\rowcolor{gray!15}
1000 & 0.737 & 1.704 & 0.967 & 1.164 & 0.190 & 3.150 & 1.618 & 3.604 & 1.986 & 2.567 & 0.265 & 6.581 \\
\hline
\end{tabular}
\end{table*}

\begin{table*}
\centering
\scriptsize
\captionof{table}{Comparison of 3.5\,GHz sampling strategies within the communicable region: MAE and RMSE.}
\label{tab:partial_mae_rmse}
\begin{tabular}{lcccccc|cccccc}
\hline
& \multicolumn{6}{c|}{\textbf{MAE}} & \multicolumn{6}{c}{\textbf{RMSE}} \\
\textbf{Case} & \textbf{Q1} & \textbf{Q3} & \textbf{IQR} & \textbf{Median} & \textbf{L-Whisker} & \textbf{U-Whisker}
& \textbf{Q1} & \textbf{Q3} & \textbf{IQR} & \textbf{Median} & \textbf{L-Whisker} & \textbf{U-Whisker} \\
\hline
 Kriging
& 2.460
& 8.537
& 6.077
& 5.773
& 0.000
& 17.243
& 4.965
& 11.601
& 6.636
& 8.679
& 0.000
& 21.424 \\
Sparse Random   & 1.912 & 4.239 & 2.327 & 3.233 & 0.652 & 7.704 & 3.645 & 7.097 & 3.451 & 5.547 & 0.861 & 12.271 \\
Sparse NLoS     & 1.934 & 4.186 & 2.252 & 3.180 & 0.647 & 7.538 & 3.634 & 6.920 & 3.286 & 5.449 & 0.884 & 11.849 \\
Blend Coverage  & 2.024 & 4.476 & 2.452 & 3.386 & 0.623 & 8.138 & 3.758 & 7.290 & 3.532 & 5.746 & 0.853 & 12.575 \\
\rowcolor{gray!15}
Block Coverage  & 1.744 & 4.025 & 2.281 & 2.975 & 0.579 & 7.432 & 3.380 & 6.796 & 3.415 & 5.308 & 0.759 & 11.904 \\
Full CommUNext  & 2.563 & 5.631 & 3.069 & 4.371 & 0.924 & 10.202 & 4.550 & 8.728 & 4.178 & 6.940 & 1.205 & 14.907 \\
\hline
\end{tabular}
\end{table*}

\subsubsection{7\,GHz Directional (Angle) Comparison}
\label{sec:Directional}
The previous results demonstrated that the model maintains directional completion capability even when the eight-direction 7\,GHz sparse SS inputs are reduced to four.
To further examine the effect of directional sparsity, we evaluate performance using fewer input directions.
Building upon the Full CommUNext architecture, five configurations are tested by varying the number of 7\,GHz sparse SS input directions:
\begin{enumerate}
    \item[i.] \textbf{4 Dir:} $0^\circ, 90^\circ, 180^\circ, 270^\circ$.
    \item[ii.] \textbf{2 Dir:} $0^\circ, 180^\circ$.
    \item[iii.] \textbf{1 Dir:} $0^\circ$.
    \item[iv.] \textbf{1 Random:} One randomly selected from the eight canonical directions.
    \item[v.] \textbf{0 Dir:} No high-frequency directional input; the model relies on the remaining inputs.
\end{enumerate}
As in the setting in Section~\ref{sec:ModelOpt}, in these experiments, each 7\,GHz sparse SS map $\widetilde{\mathbf{S}}_{di}$ contains $N_d = 200$ randomly sampled pixels.

Table~\ref{tab:dir_mae_rmse} summarizes the MAE and RMSE under different directional sparsity configurations. The results indicate that reducing the number of high-frequency directional inputs does not lead to substantial performance degradation. Although \textbf{4 Dir} provides additional directional cues, it may introduce minor instability when sparse samples are drawn from challenging regions. In contrast, \textbf{1 Random} achieves robust performance, likely because sample-wise directional diversity enriches the learned angular features. Remarkably, \textbf{0 Dir} is still able to reconstruct eight-direction 7\,GHz SS maps with reasonable accuracy. This behavior can be attributed to the fact that the eight canonical directions are fixed across all samples, which reduces the reliance on explicit directional anchors for orientation calibration.

To further quantify the difference between \textbf{4 Dir} and \textbf{0 Dir} within the communicable region, we examine their MAE and RMSE distributions. While \textbf{0 Dir} exhibits a slightly less concentrated error distribution and higher RMSE values in certain regions due to the absence of high-frequency directional references, its median MAE remains close to that of \textbf{4 Dir}. This indicates that, despite increased variance, the overall prediction accuracy of \textbf{0 Dir} remains comparable, demonstrating the robustness of the proposed framework under missing high-frequency directional inputs.

Illustrative predictions for the \textbf{0 Dir} case are shown in Fig.~\ref{fig:test_samples}(d), (j), and (p), with the corresponding absolute error maps presented in Fig.~\ref{fig:error_maps}(c), (h), and (m). Darker tones indicate lower errors in open LoS areas and along beam lobes, highlighting the framework’s cross-band and multi-directional completion capability.

\subsubsection{7\,GHz Sparse Sampling Strategy Comparison}
As discussed in the previous subsection, the proposed framework can reconstruct eight-direction SS maps even in the absence of high-frequency directional inputs. We now investigate whether alternative sampling strategies can further improve performance under a fixed total sample budget. Prior analyses indicate that major prediction errors predominantly occur in NLoS open regions. Accordingly, we adopt the \emph{NLoS-guided sampling} strategy described in Section~\ref{sec:7GHzSSMapGen}, in which each 7\,GHz sparse SS map is generated by selecting sampling points with ${\gamma = 90\%}$ drawn from NLoS open regions and the remaining $10\%$ from LoS non-building areas. The sampling density is varied as ${N_d = 200}$, $400$, $600$, $800$, and $1{,}000$ points per map.

We first compare the performance of Random Sampling and NLoS-guided sampling under the same sampling budget of $N_d = 200$ points per map across eight directions. Specifically, the \textbf{8 Dir} results in Table~\ref{tab:mae_rmse_results} are contrasted with the corresponding 200-point case in Table~\ref{tab:nloS_mae_rmse}. The results show that NLoS-guided sampling consistently yields lower MAE and RMSE values across multiple statistical indicators, reflecting both a reduction in overall error magnitude and a suppression of outliers.
Furthermore, as the number of NLoS-guided sampling points increases from 200 to 1,000 per map, both the median MAE and RMSE decrease progressively, with the best performance achieved at 1,000 points. This trend confirms that denser sampling in NLoS open regions, where prediction uncertainty is typically highest, effectively reduces large errors and enhances overall prediction accuracy.

As illustrated in Fig.~\ref{fig:test_samples}(e), (k), and (q), together with the corresponding error maps in Fig.~\ref{fig:error_maps}(d), (i), and (n), the 1,000-point NLoS-guided sampling configuration produces the most accurate predictions, with visibly reduced error concentrations.

\subsubsection{Discussion}
{
Notably, although the dataset is collected within New York City, the area is spatially partitioned into 14,884 \emph{non-overlapping} \SI{500}{m}$\times$\SI{500}{m} blocks. The training and test sets are constructed using regions with strictly disjoint building layouts, ensuring evaluation on previously \emph{unseen} urban blocks. As shown in Fig.~\ref{fig:test_samples}(a), (g), and (m), different regions exhibit markedly different SS distributions, highlighting the nontrivial nature of prediction across unseen building configurations.
}

The three sets of experiments under Full CommUNextin the previous subsections lead to the following observations:
\begin{itemize}
    \item \textbf{Cross-band and multi-directional prediction:} Leveraging complete low-frequency coverage with high-frequency sparse samples enables learning of inter-band and inter-directional correspondences.
    \item \textbf{Seg architecture advantage:} Incorporating NLoS masks with multi-task learning improves accuracy over the baseline U-Net.
    \item \textbf{No high-frequency inputs:} The model can reconstruct reasonable eight-direction SS maps without high-frequency directional data, showing generalization under unknown band and direction conditions.
    \item \textbf{Dense NLoS sampling:} Increasing sample density in NLoS open regions substantially reduces errors and improves robustness.
    \item \textbf{Role of low-frequency coverage:} Complete 3.5\,GHz maps provide essential context for reliable predictions. The \textbf{Sparse Coverage} results show that lacking full low-frequency maps degrades performance significantly, motivating Partial CommUNext for scenarios with incomplete low-frequency coverage.
\end{itemize}

\subsection{Partial CommUNext (Incomplete 3.5\,GHz Coverage Map)}

\subsubsection{3.5\,GHz Sampling Strategies and Experimental Setup}
To address realistic scenarios with incomplete low-frequency coverage, we adopt the {Partial CommUNext} architecture described in Section~\ref{sec:PartialCommUNext}, together with the 3.5\,GHz sampling strategies introduced in Section~\ref{sec:3.5GHz_Sampled_Maps}. Sparse 3.5\,GHz coverage maps are generated using $N_c = 1{,}000$ samples with $\gamma = 90\%$ under different sampling strategies and are used as training inputs.
Five configurations are considered:
\begin{enumerate}
    \item[i.] \textbf{Sparse Random:} Random sampling, identical to the sampling method used in the Sparse Coverage case of Full CommUNext.
    \item[ii.] \textbf{Sparse NLoS:} Sampling guided by NLoS regions.
    \item[iii.] \textbf{Blend Coverage:} Pseudo-point expansion from the initial samples to achieve higher sampling density.
    \item[iv.] \textbf{Block Coverage:} Block-based sampling that targets NLoS regions.
    \item[v.] \textbf{Full CommUNext:} Results from the Sparse Coverage case of Full CommUNext, included for comparison.
\end{enumerate}
All configurations are applied to the 3.5\,GHz coverage maps rather than to the 7\,GHz data. The objective is to determine which sparse 3.5\,GHz sampling strategy most effectively compensates for missing complete coverage.
For the 7\,GHz input, four fixed directional sparse SS maps at $0^\circ$, $90^\circ$, $180^\circ$, and $270^\circ$ are used, each containing $N_d = 200$ randomly sampled pixels per direction.
All Partial CommUNext experiments are trained with $\lambda_\text{Seg} = 0.3$ and $\lambda_\text{Cov} = 0.5$ in~\eqref{eq:L_partial}.

\begin{table*}[!t]
\caption{Comparison of median MAE/RMSE under four sampling scenarios within the communicable region.}
\label{tab:partial_category_mae_rmse}
\centering
\footnotesize
\setlength{\tabcolsep}{6pt}
\renewcommand{\arraystretch}{1.15}
\begin{tabular}{l|cc|cc|cc|cc}
\hline
& \multicolumn{2}{c|}{\textbf{3.5\,GHz \& 7\,GHz Sampled}}
& \multicolumn{2}{c|}{\textbf{Only 3.5\,GHz Sampled}}
& \multicolumn{2}{c|}{\textbf{Only 7\,GHz Sampled}}
& \multicolumn{2}{c}{\textbf{Neither Sampled}} \\
\textbf{Case}
& \textbf{MAE$_\text{med}$} & \textbf{RMSE$_\text{med}$}
& \textbf{MAE$_\text{med}$} & \textbf{RMSE$_\text{med}$}
& \textbf{MAE$_\text{med}$} & \textbf{RMSE$_\text{med}$}
& \textbf{MAE$_\text{med}$} & \textbf{RMSE$_\text{med}$} \\
\hline
Sparse Random   & 1.709 & 2.055 & 2.067 & 3.354 & 2.016 & 2.867 & 3.348 & 5.614 \\
Sparse NLoS     & 1.602 & 1.941 & 1.947 & 3.123 & 1.884 & 2.632 & 3.263 & 5.574 \\
Blend Coverage  & 2.149 & 2.962 & 2.958 & 4.735 & 2.163 & 2.989 & 3.813 & 6.400 \\
Block Coverage  & 1.582 & 2.031 & 1.995 & 3.506 & 1.849 & 2.627 & 3.181 & 5.531 \\
Full CommUNext  & 2.112 & 2.602 & 4.327 & 6.818 & 2.445 & 3.404 & 4.372 & 6.941 \\
\hline
\end{tabular}
\end{table*}

\subsubsection{Performance Comparison of Sampling Strategies}
Table~\ref{tab:partial_mae_rmse} compares the \textbf{Sparse Random} case with the Kriging and Full CommUNext baselines, where both settings use identical randomly sampled 3.5\,GHz sparse maps. The results show that Partial CommUNext achieves superior prediction accuracy under this setting. Moreover, \textbf{Sparse NLoS} consistently outperforms \textbf{Sparse Random}, confirming that directing samples toward challenging NLoS regions provides more informative guidance for high-frequency prediction. In contrast, \textbf{Blend Coverage}, despite increasing the number of pseudo-points, introduces cumulative errors that degrade performance relative to direct sparse sampling. By comparison, \textbf{Block Coverage} achieves the best overall performance by balancing focused sampling in difficult NLoS areas with the preservation of local structural features, thereby enabling finer reconstruction of propagation characteristics.
Overall, although Partial CommUNext does not reach the accuracy of Full CommUNext with complete 3.5\,GHz coverage, it consistently outperforms Full CommUNext when only sparsely sampled low-frequency maps are available.

Further insights can be obtained from the MAE and RMSE statistics in Table~\ref{tab:partial_mae_rmse}. While the MAE differences among the four Partial CommUNext configurations are relatively small, the RMSE upper whiskers for \textbf{Sparse NLoS} and \textbf{Block Coverage} are notably lower than those for \textbf{Sparse Random} and \textbf{Blend Coverage}. This indicates that sampling more points from NLoS and other challenging regions not only reduces overall errors but also effectively suppresses large outliers. In contrast, although \textbf{Blend Coverage} also concentrates samples in NLoS regions, the replicated pseudo-points may mislead the model and result in larger prediction errors.

Qualitative results are shown in Fig.~\ref{fig:test_samples}(f), (l), and (r), which present predictions obtained using the 3.5\,GHz Block Coverage strategy. Compared with the \textbf{Sparse Coverage} case in Fig.~\ref{fig:test_samples}(c), (i), and (o), block-based sampling better preserves ray-tracing path structures. For example, the upper-right region in Fig.~\ref{fig:test_samples}(f) exhibits improved reconstruction, indicating enhanced retention of local features. The corresponding error maps in Fig.~\ref{fig:error_maps}(e), (j), and (o) further confirm reduced errors in NLoS open regions compared with Full CommUNext under the \textbf{Sparse Coverage} setting shown in Fig.~\ref{fig:error_maps}(b), (g), and (l).

\subsubsection{Impact of Input Availability}
To analyze the importance of different input sources, Table~\ref{tab:partial_category_mae_rmse} further refines the results in Table~\ref{tab:partial_mae_rmse} by categorizing prediction performance according to whether a pixel is assisted by a 3.5\,GHz sampled point, a 7\,GHz sparse SS point from other directions, both, or neither. A consistent trend is observed. Configurations that combine 7\,GHz sparse SS samples with 3.5\,GHz sampled points achieve the highest accuracy, followed by those that rely solely on 7\,GHz sparse SS points. Notably, even a single 7\,GHz sparse SS point provides directional cues that facilitate the reconstruction of SS values across all other directions at 7\,GHz.

Using only 3.5\,GHz sampled points ranks third but still yields noticeable error reduction through cross-band guidance. When comparing the cases that use only 7\,GHz inputs with those that use only 3.5\,GHz inputs, it is evident that a 7\,GHz sparse SS point provides richer information, as it lies in the target frequency band, even when only a single directional cue is available. As expected, the absence of both reference sources results in the worst performance, where predictions rely primarily on spatial correlations inferred from nearby users. Notably, the median MAE and RMSE values of the \emph{Neither Sampled} category are close to, but slightly worse than, the aggregate results in Table~\ref{tab:partial_mae_rmse}. This behavior arises because most pixels fall into the \emph{Neither Sampled} category, while the relatively few pixels with 3.5\,GHz or 7\,GHz samples contribute only marginal improvements, leading to the observed discrepancy.

\begin{figure} 
\centering
{
\includegraphics[width=0.45\textwidth]{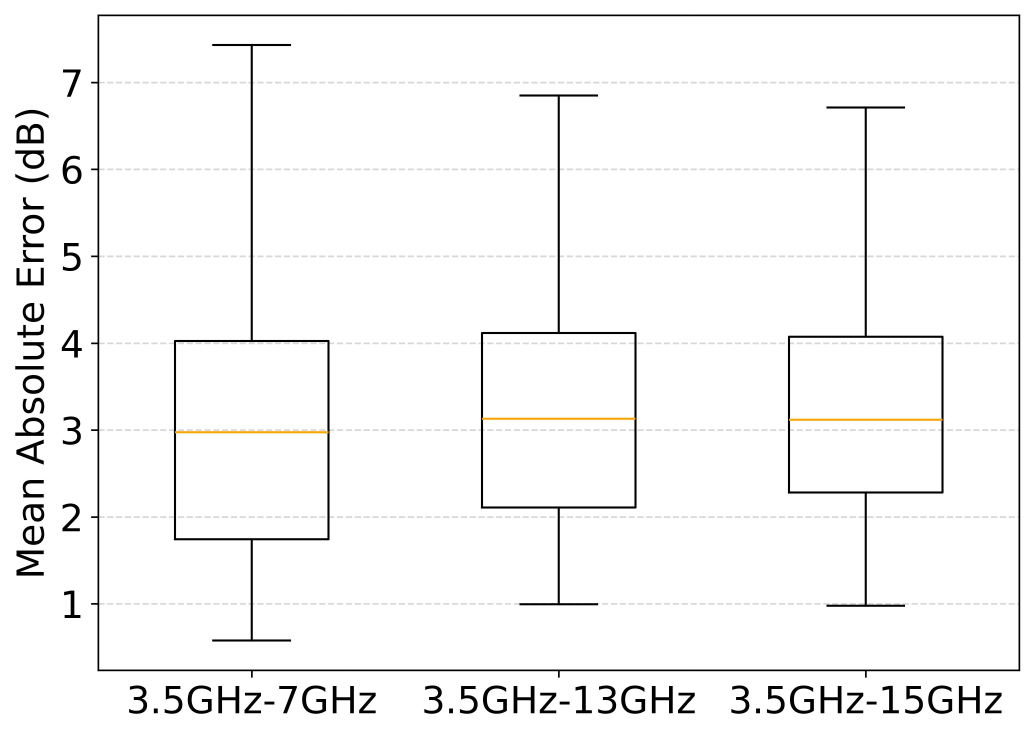}
\caption{Box plots of MAE across different FR1 and FR3 frequency pairs.}
\label{fig:diffpairs_mae}
}
\end{figure}

\subsubsection{Cross-Band Feasibility Across FR1 and FR3 Bands}
{
We further evaluate the cross-band feasibility of Partial CommUNext across different FR1 and FR3 frequency pairs. In addition to the 3.5\,GHz and 7\,GHz setting, the 3.5\,GHz and 13\,GHz and the 3.5\,GHz and 15\,GHz pairs are also considered. For each frequency pair, the model is fine-tuned for only $50$ epochs using the Block Coverage sampling strategy. This lightweight fine-tuning primarily adapts frequency-dependent attenuation characteristics while reusing the learned geometric representations, thereby enabling stable performance across different frequency pairs. Fig.~\ref{fig:diffpairs_mae} reports the MAE box plots within the communicable region, which is consistently defined based on the 7\,GHz coverage map to ensure fair comparison across frequency pairs. The comparable error levels across target frequencies indicate that the proposed framework maintains stable performance when extended to higher FR3 bands with limited fine-tuning.

Beyond the comparable median MAE values, an interesting observation  can be drawn from Fig.~\ref{fig:diffpairs_mae}. As the target frequency increases from 7\,GHz to 13\,GHz and 15\,GHz, the median error exhibits only marginal variation, while the IQR slightly decreases, indicating a more concentrated error distribution. One possible explanation is that, at higher FR3 frequencies, the number of viable propagation paths in NLoS regions is reduced due to increased attenuation of higher-order reflections and diffractions. Consequently, the effective multipath structure becomes sparser and more strongly dominated by a limited number of primary paths. When combined with low-frequency coverage and partial high-frequency observations, this reduced path diversity may lower the conditional uncertainty in NLoS regions, resulting in more stable predictions.
}

\subsubsection{Discussion}
Several conclusions can be drawn from the preceding experiments:
\begin{itemize}
    \item \textbf{Partial CommUNext vs. Full CommUNext:} Under identical sampling maps, Partial CommUNext achieves superior prediction compared with Full CommUNext, as evidenced by the \textbf{Sparse Random} case.

    \item \textbf{NLoS-guided sampling improves accuracy:} \textbf{Sparse NLoS} outperforms \textbf{Sparse Random}, confirming that sampling difficult NLoS regions provides valuable guidance for high-frequency prediction.

    \item \textbf{Block-based sampling balances difficulty and structure:} \textbf{Block Coverage} yields the best results, capturing NLoS areas while preserving structural features and enabling finer reconstruction of local propagation characteristics.

    \item \textbf{Practical potential of Partial CommUNext:} While not as precise as Full CommUNext with full coverage, Partial CommUNext combined with appropriate sampling strategies can reconstruct 3.5\,GHz maps and predict complete eight-direction 7\,GHz SS maps, offering strong applicability to real-world scenarios where full low-frequency coverage is unavailable.
\end{itemize}

We conclude this section by discussing computational efficiency.
Compared with RT-based simulations, CommUNext substantially reduces inference cost. On an RTX~2080~Ti GPU, generating eight directional SS maps requires approximately $0.002$\,s for Full CommUNext and $0.01$\,s for Partial CommUNext, whereas RT requires about $19$\,s per map, corresponding to roughly $152$\,s for eight maps.
The number of trainable parameters can be expressed as $\text{Params} = P_0 + {576\times C_{\mathrm{in}}} + {65\times C_{\mathrm{out}}}$, where $C_{\mathrm{in}}$ and $C_{\mathrm{out}}$ denote the numbers of input and output channels, respectively. Here, $P_0 = 31{,}036{,}096$ for Full CommUNext and $P_0 = 44{,}277{,}122$ for Partial CommUNext.

\section{Conclusion}
This study proposed CommUNext, a deep learning framework for cross-band and multi-directional SS prediction, and validated its performance under different levels of data completeness. Experimental results show that, when a complete low-frequency 3.5~GHz coverage map is available, Full CommUNext effectively enables accurate prediction of multi-directional high-frequency 7~GHz SS maps. Even in scenarios with sparse high-frequency directional inputs or incomplete low-frequency information, the proposed framework maintains reasonable prediction performance. To address the practical challenge of acquiring complete low-frequency coverage maps, the Partial CommUNext architecture was further developed. By leveraging carefully designed sampling strategies, Partial CommUNext reconstructs low-frequency coverage maps from limited 3.5~GHz samples and subsequently predicts the corresponding high-frequency SS distributions.

Overall, this work demonstrates that efficient and robust communication prediction can be achieved under constrained data conditions through appropriate architectural design and sampling strategies. Nevertheless, the current analysis assumes co-located FR1 and FR3 sites with limited antenna displacement, under which spatial correlation across frequency bands remains strong. Future work will focus on validating the proposed framework using real-world measurement data and extending it to scenarios involving significant antenna reorientation or non-co-located deployments, where cross-band spatial correlation is weakened.

\appendices

\section{Operational Assumptions for Crowd-Aided Measurements} \label{sec:appendix_a}
The crowd-aided operational scenario considered in this work is defined under a set of explicit and controlled assumptions, which are clarified in this appendix to avoid ambiguity.

First, we assume that sparse measurements within a local spatial region are collected using the same mobile device model. Under this assumption, variations in the reported SS primarily arise from user handling and device orientation, which induce orientation-dependent antenna gain fluctuations. Rather than explicitly modeling the temporal measurement process, we operate at an aggregated representation level, where such orientation-induced variations are treated as stochastic effects and implicitly marginalized. As a result, the SS associated with each location is modeled as an orientation-averaged quantity.

When measurements originate from different device models, the dominant discrepancy can be reasonably approximated as a device-dependent SS offset, mainly caused by differences in antenna radiation patterns and front-end calibration. Such offsets can be mitigated through normalization or calibration procedures and do not alter the underlying spatial or directional propagation characteristics. Consequently, the proposed framework focuses on learning spatial, directional, and cross-band relationships, while device-specific gain biases are not explicitly modeled.

Based on these assumptions, the synthetic sampling strategies adopted in this work are designed to emulate the spatial sparsity, directional incompleteness, and aggregation behavior inherent in crowd-aided measurement reports.

\section{Channel Characterization and Relevance to FR1-FR3 Prediction}
\label{app:channel_rmap}
This appendix briefly clarifies the radio map-based channel quantity considered in this work and its relevance to the FR1-FR3 cross band prediction setting. The ray tracing formulation follows the Sionna RT technical report~\cite{aoudia2025sionnarttr}.

For a receiver location $\mathbf{x}$ on the measurement surface $\mathcal{M}$, ray tracing produces a set of propagation paths $\mathcal{P}_\mathbf{x}$ reaching $\mathbf{x}$. According to~\cite[Eq. (5)]{aoudia2025sionnarttr}, the radio map value at $\mathbf{x}$ is given by
\begin{equation} \label{eq:R_x}
R_\mathbf{x} = \sum_{p \in \mathcal{P}_\mathbf{x}} G(p, f_c),
\end{equation}
where $f_c$ is the carrier frequency and $G(p,f_c)$ denotes the path gain of path $p$, including free space propagation and interaction losses. In this work, the wireless channel is represented at the radio map level, and each pixel value of the radio map corresponds to the total received path gain in \eqref{eq:R_x}.

The radio map is determined by the environment geometry and frequency dependent propagation effects. While the geometry constrains the propagation path structure and is independent of frequency, frequency dependent reflection and scattering losses lead to different radio map values across bands. Rather than relying on explicit model-based per path transformations across frequencies \cite{CrossBandCIR2025}, which require accurate characterization of frequency dependent interaction losses and path-wise correspondence, a learning-based approach is adopted to learn the cross band mapping directly from Sionna RT radio maps as supervision.

\bibliographystyle{IEEEtran}

\end{document}